\documentclass[twocolumn]{aastex631}
\reportnum{FERMILAB-PUB-25-0038}
\reportnum{DES-2024-0875}

\usepackage{import}
\usepackage{booktabs}
\usepackage{graphicx}
\usepackage{tabularx}
\usepackage{longtable}
\usepackage{ltablex}
\usepackage{threeparttable}
\let\tablenum\relax
\usepackage[detect-all]{siunitx}

\begin{document}

\title{Discovering Strong Gravitational Lenses in the Dark Energy Survey with Interactive Machine Learning and Crowd-sourced Inspection with Space Warps}

\author[0000-0001-7282-3864]{J. González}\affiliation{Physics Department, University of Wisconsin-Madison, 1150 University Avenue Madison, WI 53706, USA \\ \href{mailto:sgonzalezloz@wisc.edu}{sgonzalezloz@wisc.edu}}
\author[0009-0002-8896-6100]{P. Holloway}\affiliation{Sub-department of Astrophysics, University of Oxford, Denys Wilkinson Building, Keble Road, Oxford OX1 3RH, UK}
\author[0000-0001-5564-3140]{T. Collett}\affiliation{Institute of Cosmology and Gravitation, University of Portsmouth, Burnaby Rd, Portsmouth PO1 3FX, UK}
\author[0000-0002-0730-0781]{A. Verma}\affiliation{Sub-department of Astrophysics, University of Oxford, Denys Wilkinson Building, Keble Road, Oxford OX1 3RH, UK}
\author[0000-0001-8156-0429]{K. Bechtol}\affiliation{Physics Department, University of Wisconsin-Madison, 1150 University Avenue Madison, WI 53706, USA}
\author[0000-0002-0113-5770]{P. Marshall}\affiliation{Kavli Institute for Particle Astrophysics and Cosmology, Department of Physics, Stanford University, Stanford, CA 94305, USA} \affiliation{SLAC National Accelerator Laboratory, Menlo Park, CA 94025, USA}
\author[0000-0001-7714-7076]{A. More}\affiliation{The Inter-University Centre for Astronomy and Astrophysics (IUCAA), Post Bag 4, Ganeshkhind, Pune 411007, India}\affiliation{Kavli Institute for the Physics and Mathematics of the Universe (IPMU), 5-1-5 Kashiwanoha, Kashiwa-shi, Chiba 277-8583, Japan}
\author[0000-0002-9654-1711]{J. Acevedo Barroso}\affiliation{Institute of Physics, Laboratory of Astrophysics, Ecole Polytechnique F\'ed\'erale de Lausanne (EPFL), Observatoire de Sauverny, 1290 Versoix, Switzerland}
\author[0009-0005-7030-9948]{G. Cartwright}\affiliation{Physics Department, University of Wisconsin-Madison, 1150 University Avenue Madison, WI 53706, USA} \affiliation{School of Physics and Astronomy, University of Minnesota, 116 Church Street S.E., Minneapolis, MN 55455, USA}
\author[0000-0002-8397-8412]{M. Martinez}\affiliation{Physics Department, University of Wisconsin-Madison, 1150 University Avenue Madison, WI 53706, USA}
\author[0009-0005-5008-0381]{T. Li} \affiliation{Institute of Cosmology and Gravitation, University of Portsmouth, Burnaby Rd, Portsmouth PO1 3FX, UK}
\author[0000-0003-1391-6854]{K. Rojas} \affiliation{Institute of Cosmology and Gravitation, University of Portsmouth, Burnaby Rd, Portsmouth PO1 3FX, UK} 
\author[0000-0003-2497-6334]{S. Schuldt} \affiliation{Dipartimento di Fisica, Università degli Studi di Milano, via Celoria 16, I-20133 Milano, Italy
and: INAF–IASF Milano, via A. Corti 12, I-20133 Milano, Italy}
\author[0000-0003-3195-5507]{S. Birrer} \affiliation{Department of Physics and Astronomy, Stony Brook University, Stony Brook, NY 11794, USA}
\author[0000-0002-8357-7467]{H. T. Diehl} \affiliation{Fermi National Accelerator Laboratory, P. O. Box 500, Batavia, IL 60510, USA}
\author[0000-0002-7016-5471]{R. Morgan}\affiliation{Physics Department, University of Wisconsin-Madison, 1150 University Avenue Madison, WI 53706, USA}
\author[0000-0001-8251-933X]{A. Drlica-Wagner} \affiliation{Fermi National Accelerator Laboratory, P.O.\ Box 500, Batavia, IL 60510, USA} \affiliation{Kavli Institute for Cosmological Physics, University of Chicago, Chicago, IL 60637, USA} \affiliation{Department of Astronomy and Astrophysics, University of Chicago, Chicago, IL 60637, USA}
\author[0000-0003-4083-1530]{J. H. O'Donnell} \affiliation{Department of Physics, University of California, Santa Cruz (UCSC), Santa Cruz, CA 95064,
USA} \affiliation{Santa Cruz Institute for Particle Physics (SCIPP), Santa Cruz, CA 95064, USA}
\author[0000-0002-6779-4277]{E. Zaborowski} \affiliation{Department of Physics, The Ohio State University, 191 West Woodruff Avenue, Columbus, OH 43210, USA} \affiliation{Center for Cosmology and AstroParticle Physics, The Ohio State University, 191 West Woodruff Avenue, Columbus, OH 43210, USA}
\author[0000-0001-6706-8972]{B. Nord} \affiliation{Fermi National Accelerator Laboratory, P.O. Box 500, Batavia, Illinois 60510, USA} \affiliation{Department of Astronomy and Astrophysics, University of Chicago, Chicago, Illinois 60637, USA} \affiliation{Kavli Institute for Cosmological Physics, University of Chicago, Chicago, Illinois 60637, USA}
\author{E. M. Baeten} \affiliation{Citizen Scientist, Zooniverse c/o University of Oxford, Keble Road, Oxford OX1 3RH, UK}
\author[0000-0001-6421-0953]{L. C. Johnson} \affiliation{Center for Interdisciplinary Exploration and Research in Astrophysics (CIERA) and Department of Physics and Astronomy, Northwestern University, 1800 Sherman Avenue, Evanston, IL 60201, USA}
\author{C. Macmillan} \affiliation{Zooniverse, c/o Oxford Astrophysics, Denys Wilkinson Building, Keble Road, Oxford OX1 3RH, UK}
\author{T.~M.~C.~Abbott} \affiliation{Cerro Tololo Inter-American Observatory, NSF's National Optical-Infrared Astronomy Research Laboratory, Casilla 603, La Serena, Chile}
\author{M.~Aguena} \affiliation{Laborat\'orio Interinstitucional de e-Astronomia - LIneA, Rua Gal. Jos\'e Cristino 77, Rio de Janeiro, RJ - 20921-400, Brazil}
\author[0000-0002-7069-7857]{S.~S.~Allam} \affiliation{Fermi National Accelerator Laboratory, P. O. Box 500, Batavia, IL 60510, USA}
\author[0000-0002-8458-5047]{D.~Brooks} \affiliation{Department of Physics \& Astronomy, University College London, Gower Street, London, WC1E 6BT, UK}
\author[0000-0002-3304-0733]{E.~Buckley-Geer} \affiliation{Department of Astronomy and Astrophysics, University of Chicago, Chicago, IL 60637, USA} \affiliation{Fermi National Accelerator Laboratory, P. O. Box 500, Batavia, IL 60510, USA}
\author{D.~L.~Burke} \affiliation{Kavli Institute for Particle Astrophysics \& Cosmology, P. O. Box 2450, Stanford University, Stanford, CA 94305, USA} \affiliation{SLAC National Accelerator Laboratory, Menlo Park, CA 94025, USA}
\author[0000-0003-3044-5150]{A.~Carnero~Rosell} \affiliation{Instituto de Astrofisica de Canarias, E-38205 La Laguna, Tenerife, Spain} \affiliation{Laborat\'orio Interinstitucional de e-Astronomia - LIneA, Rua Gal. Jos\'e Cristino 77, Rio de Janeiro, RJ - 20921-400, Brazil} \affiliation{Universidad de La Laguna, Dpto. Astrofísica, E-38206 La Laguna, Tenerife, Spain}
\author[0000-0002-3130-0204]{J.~Carretero} \affiliation{Institut de F\'{\i}sica d'Altes Energies (IFAE), The Barcelona Institute of Science and Technology, Campus UAB, 08193 Bellaterra (Barcelona) Spain}
\author{R.~Cawthon} \affiliation{Physics Department, William Jewell College, Liberty, MO, 64068}
\author[0000-0002-4213-8783]{T.~M.~Davis} \affiliation{School of Mathematics and Physics, University of Queensland,  Brisbane, QLD 4072, Australia}
\author[0000-0001-8318-6813]{J.~De~Vicente} \affiliation{Centro de Investigaciones Energ\'eticas, Medioambientales y Tecnol\'ogicas (CIEMAT), Madrid, Spain}
\author[0000-0002-0466-3288]{S.~Desai} \affiliation{Department of Physics, IIT Hyderabad, Kandi, Telangana 502285, India}
\author{P.~Doel} \affiliation{Department of Physics \& Astronomy, University College London, Gower Street, London, WC1E 6BT, UK}
\author{S.~Everett} \affiliation{California Institute of Technology, 1200 East California Blvd, MC 249-17, Pasadena, CA 91125, USA}
\author[0000-0002-2367-5049]{B.~Flaugher} \affiliation{Fermi National Accelerator Laboratory, P. O. Box 500, Batavia, IL 60510, USA}
\author[0000-0003-4079-3263]{J.~Frieman} \affiliation{Fermi National Accelerator Laboratory, P. O. Box 500, Batavia, IL 60510, USA} \affiliation{Kavli Institute for Cosmological Physics, University of Chicago, Chicago, IL 60637, USA}
\author[0000-0002-9370-8360]{J.~Garc\'ia-Bellido} \affiliation{Instituto de Fisica Teorica UAM/CSIC, Universidad Autonoma de Madrid, 28049 Madrid, Spain}
\author[0000-0001-9632-0815]{E.~Gaztanaga} \affiliation{Institut d'Estudis Espacials de Catalunya (IEEC), 08034 Barcelona, Spain} \affiliation{Institute of Cosmology and Gravitation, University of Portsmouth, Portsmouth, PO1 3FX, UK} \affiliation{Institute of Space Sciences (ICE, CSIC),  Campus UAB, Carrer de Can Magrans, s/n,  08193 Barcelona, Spain}
\author[0000-0002-3730-1750]{G.~Giannini} \affiliation{Institut de F\'{\i}sica d'Altes Energies (IFAE), The Barcelona Institute of Science and Technology, Campus UAB, 08193 Bellaterra (Barcelona) Spain} \affiliation{Kavli Institute for Cosmological Physics, University of Chicago, Chicago, IL 60637, USA}
\author[0000-0003-3270-7644]{D.~Gruen} \affiliation{University Observatory, Faculty of Physics, Ludwig-Maximilians-Universit\"at, Scheinerstr. 1, 81679 Munich, Germany}
\author{R.~A.~Gruendl} \affiliation{Center for Astrophysical Surveys, National Center for Supercomputing Applications, 1205 West Clark St., Urbana, IL 61801, USA} \affiliation{Department of Astronomy, University of Illinois at Urbana-Champaign, 1002 W. Green Street, Urbana, IL 61801, USA}
\author[0000-0003-0825-0517]{G.~Gutierrez} \affiliation{Fermi National Accelerator Laboratory, P. O. Box 500, Batavia, IL 60510, USA}
\author{S.~R.~Hinton} \affiliation{School of Mathematics and Physics, University of Queensland,  Brisbane, QLD 4072, Australia}
\author{D.~L.~Hollowood} \affiliation{Santa Cruz Institute for Particle Physics, Santa Cruz, CA 95064, USA}
\author[0000-0002-6550-2023]{K.~Honscheid} \affiliation{Center for Cosmology and Astro-Particle Physics, The Ohio State University, Columbus, OH 43210, USA} \affiliation{Department of Physics, The Ohio State University, Columbus, OH 43210, USA}
\author[0000-0001-5160-4486]{D.~J.~James} \affiliation{Center for Astrophysics $\vert$ Harvard \& Smithsonian, 60 Garden Street, Cambridge, MA 02138, USA}
\author[0000-0003-0120-0808]{K.~Kuehn} \affiliation{Australian Astronomical Optics, Macquarie University, North Ryde, NSW 2113, Australia} \affiliation{Lowell Observatory, 1400 Mars Hill Rd, Flagstaff, AZ 86001, USA}
\author[0000-0002-1134-9035]{O.~Lahav} \affiliation{Department of Physics \& Astronomy, University College London, Gower Street, London, WC1E 6BT, UK}
\author{S.~Lee} \affiliation{Jet Propulsion Laboratory, California Institute of Technology, 4800 Oak Grove Dr., Pasadena, CA 91109, USA}
\author{M.~Lima} \affiliation{Departamento de F\'isica Matem\'atica, Instituto de F\'isica, Universidade de S\~ao Paulo, CP 66318, S\~ao Paulo, SP, 05314-970, Brazil} \affiliation{Laborat\'orio Interinstitucional de e-Astronomia - LIneA, Rua Gal. Jos\'e Cristino 77, Rio de Janeiro, RJ - 20921-400, Brazil}
\author[0000-0003-0710-9474]{J.~L.~Marshall} \affiliation{George P. and Cynthia Woods Mitchell Institute for Fundamental Physics and Astronomy, and Department of Physics and Astronomy, Texas A\&M University, College Station, TX 77843,  USA}
\author[0000-0001-9497-7266]{J. Mena-Fern{\'a}ndez} \affiliation{LPSC Grenoble - 53, Avenue des Martyrs 38026 Grenoble, France}
\author[0000-0002-6610-4836]{R.~Miquel} \affiliation{Instituci\'o Catalana de Recerca i Estudis Avan\c{c}ats, E-08010 Barcelona, Spain} \affiliation{Institut de F\'{\i}sica d'Altes Energies (IFAE), The Barcelona Institute of Science and Technology, Campus UAB, 08193 Bellaterra (Barcelona) Spain}
\author{J.~Myles} \affiliation{Department of Astrophysical Sciences, Princeton University, Peyton Hall, Princeton, NJ 08544, USA}
\author{M.~E.~S.~Pereira} \affiliation{Hamburger Sternwarte, Universit\"{a}t Hamburg, Gojenbergsweg 112, 21029 Hamburg, Germany}
\author[0000-0001-9186-6042]{A.~Pieres} \affiliation{Laborat\'orio Interinstitucional de e-Astronomia - LIneA, Rua Gal. Jos\'e Cristino 77, Rio de Janeiro, RJ - 20921-400, Brazil} \affiliation{Observat\'orio Nacional, Rua Gal. Jos\'e Cristino 77, Rio de Janeiro, RJ - 20921-400, Brazil}
\author[0000-0002-2598-0514]{A.~A.~Plazas~Malag\'on} \affiliation{Kavli Institute for Particle Astrophysics \& Cosmology, P. O. Box 2450, Stanford University, Stanford, CA 94305, USA} \affiliation{SLAC National Accelerator Laboratory, Menlo Park, CA 94025, USA}
\author[0000-0001-5326-3486]{A.~Roodman} \affiliation{Kavli Institute for Particle Astrophysics \& Cosmology, P. O. Box 2450, Stanford University, Stanford, CA 94305, USA} \affiliation{SLAC National Accelerator Laboratory, Menlo Park, CA 94025, USA}
\author{S.~Samuroff} \affiliation{Department of Physics, Northeastern University, Boston, MA 02115, USA} \affiliation{Institut de F\'{\i}sica d'Altes Energies (IFAE), The Barcelona Institute of Science and Technology, Campus UAB, 08193 Bellaterra (Barcelona) Spain}
\author[0000-0002-9646-8198]{E.~Sanchez} \affiliation{Centro de Investigaciones Energ\'eticas, Medioambientales y Tecnol\'ogicas (CIEMAT), Madrid, Spain}
\author[0000-0003-3054-7907]{D.~Sanchez Cid} \affiliation{Centro de Investigaciones Energ\'eticas, Medioambientales y Tecnol\'ogicas (CIEMAT), Madrid, Spain}
\author{B.~Santiago} \affiliation{Instituto de F\'\i sica, UFRGS, Caixa Postal 15051, Porto Alegre, RS - 91501-970, Brazil} \affiliation{Laborat\'orio Interinstitucional de e-Astronomia - LIneA, Rua Gal. Jos\'e Cristino 77, Rio de Janeiro, RJ - 20921-400, Brazil}
\author[0000-0002-1831-1953]{I.~Sevilla-Noarbe} \affiliation{Centro de Investigaciones Energ\'eticas, Medioambientales y Tecnol\'ogicas (CIEMAT), Madrid, Spain}
\author[0000-0002-3321-1432]{M.~Smith} \affiliation{Physics Department, Lancaster University, Lancaster, LA1 4YB, UK}
\author[0000-0002-7047-9358]{E.~Suchyta} \affiliation{Computer Science and Mathematics Division, Oak Ridge National Laboratory, Oak Ridge, TN 37831}
\author[0000-0003-1704-0781]{G.~Tarle} \affiliation{Department of Physics, University of Michigan, Ann Arbor, MI 48109, USA}
\author[0000-0001-7211-5729]{D.~L.~Tucker} \affiliation{Fermi National Accelerator Laboratory, P. O. Box 500, Batavia, IL 60510, USA}
\author{V.~Vikram} \affiliation{High Energy Physics Division, Argonne National Laboratory, 9700 S. Cass Avenue, Argonne, IL 60439, USA}
\author[0000-0002-7123-8943]{A.~R.~Walker} \affiliation{Cerro Tololo Inter-American Observatory, NSF's National Optical-Infrared Astronomy Research Laboratory, Casilla 603, La Serena, Chile}
\author{N.~Weaverdyck} \affiliation{Department of Astronomy, University of California, Berkeley,  501 Campbell Hall, Berkeley, CA 94720, USA} \affiliation{Lawrence Berkeley National Laboratory, 1 Cyclotron Road, Berkeley, CA 94720, USA}

\author{(DES Collaboration)}





\begin{abstract}
We conduct a search for strong gravitational lenses in the Dark Energy Survey (DES) Year 6 imaging data. We implement a pre-trained Vision Transformer (ViT) for our machine learning (ML) architecture and adopt Interactive Machine Learning to construct a training sample with multiple classes to address common types of false positives. Our ML model reduces \raisebox{0.5ex}{\texttildelow}236 million DES cutout images to 22,564 targets of interest, including \raisebox{0.5ex}{\texttildelow}85\% of previously reported galaxy-galaxy lens candidates discovered in DES. These targets were visually inspected by citizen scientists, who ruled out \raisebox{0.5ex}{\texttildelow}90\% as false positives. Of the remaining 2,618 candidates, 149 were expert-classified as ‘definite’ lenses and 516 as ‘probable’ lenses, with 147 of these candidates being newly identified. Additionally, we trained a second ViT to find double-source plane lens systems, finding at least one double-source system. Our main ViT excels at identifying galaxy-galaxy lenses, consistently assigning high scores to candidates with high confidence. The top 800 ViT-scored images include \raisebox{0.5ex}{\texttildelow}100 of our `definite' lens candidates. This selection is an order of magnitude higher in purity than previous convolutional neural network-based lens searches and demonstrates the feasibility of applying our methodology for discovering large samples of lenses in future surveys.
\end{abstract}



\section{Introduction} \label{sec:intro}

Gravitational lensing occurs when an astronomical source is located behind a massive foreground object such as a galaxy, galaxy group, or cluster of galaxies. If the source is sufficiently close to the line of sight between the observer and the lensing object, its light can be deflected by the gravitational potential of the foreground object, producing multiple magnified images of the source. If the source is a galaxy these images appear as extended arcs, whereas if the source is intrinsically small, such as a quasar, we see multiple magnified point source images. 

In galaxy-scale strong gravitational lenses, systems in which the deflector is a single galaxy, the observables are primarily sensitive to the mass distribution of the lensing galaxy. Most massive galaxies in the universe are elliptical galaxies, so strong lensing can be utilized to study the structure (baryonic and dark matter profiles) and evolution of these galaxies \citep{shajib2024stronglensinggalaxies}. Strong gravitational lensing is also sensitive to the underlying cosmological parameters of our Universe \citep{pascale2024snh0pemeasurementh0, Kelly_2023, Birrer_2020_tdcosmo, shajib_strides, Wong_2019}. A population of \raisebox{0.5ex}{\texttildelow}10,000 galaxy-galaxy gravitational lenses can provide competitive constraints on the dark energy equation of state parameters \citep{li2023cosmologylargepopulationsgalaxygalaxy}. Strong gravitational lenses with two sources at different redshifts (double-source plane lenses, hereafter DSPLs) are particularly useful for cosmology \citep{Linder_2016, Collett_2012} since the ratio of Einstein radii is only mildly sensitive to the mass profile of the deflector, the dominant systematic in strong lensing analyses. With only one system \cite{Collett_2014} improved the equation of state of dark energy $w$ from Planck by 30\% . 

The initial discoveries of strong gravitational lenses were mostly serendipitous (e.g., \cite{Walsh1979}). The first dedicated searches for strong lenses involved manually inspecting large numbers of images from small-area surveys applying modest selection criteria, resulting in the identification of hundreds of lenses (e.g., \cite{More_2011, Newton_2009, Faure_2008, Jackson_2008, Moustakas_2007, Fassnacht_2006, Fassnacht_2004}). However, this method is impractical for wide-field surveys, necessitating the development of autom
ated detection techniques. Early automated methods primarily targeted lenses with multiple well-separated images or distinctive elongated and curved arcs (e.g., \cite{Gavazzi_2014, More_2012, alard2006automateddetectiongravitationalarcs}). Since then, additional automated search techniques have been developed. For instance, algorithms such as `Blue Near Anything' and `Red Near Anything' focus on identifying strong lensing systems with lensed sources of the corresponding color \citep{2022ApJS..259...27O, 2017ApJS..232...15D}. While these automatic techniques have proven effective, they still require substantial visual inspection to confirm candidates. Other strong lensing search methods leverage crowdsourced visual inspection of large volumes of images, resulting in candidate samples with high purity and completeness \citep{marshall_2016_sw, more_2016_sw}.

Current state-of-the-art search methods involve the application of machine learning (ML) techniques, which have been successfully used in various fields of astrophysics \citep{Huertas_Company_2023}. Among these, Convolutional Neural Networks (CNNs), a class of deep learning algorithms suited to identify patterns and features in images, have shown significant promise. \cite{Metcalf_2019} compared different strong lensing search methodologies and found that CNN-based techniques consistently outperform previous methods, significantly reducing the presence of false positives. Numerous studies have applied these techniques to wide-field imaging surveys, identifying thousands of strong lenses (e.g., \cite{storfer2023newstronggravitationallenses, Zaborowski_2023, 2022AA...668A..73R, 2022AA...662A...4S, 2021ApJ...909...27H, canameras_2020, 2020ApJ...894...78H, 2020ApJ...899...30L, Jacobs_a, Jacobs_b, 2019MNRAS.484.3879P}). However, despite reaching excellent performance on developed testing samples, the results on real data are very different, usually exhibiting low true-positive rates of \raisebox{0.5ex}{\texttildelow}1\% or less. This discrepancy may stem from training samples that lack the realism and diversity of actual data.

Recently, a novel machine-learning technique known as the transformer encoder \citep{vaswani2023attentionneed} has been adapted to image classification tasks. This technique has shown superior performance compared to traditional CNN techniques on many image classification datasets like ImageNet, CIFAR-10 and CIFAR-100 \citep{vision_transformer}. \cite{Thuruthipilly_2024} integrated a transformer encoder in their strong lens search methodology and applied it to data from the Kilo-Degree Survey (KiDS), reporting a true-positive rate of 0.5\%. \cite{Grespan_2024} adopted the ML architectures of the previous work and fine-tuned them to real KiDS data, after applying them to \raisebox{0.5ex}{\texttildelow}5 million galaxies from KiDS, they select a sample of \raisebox{0.5ex}{\texttildelow}51,000 systems for visual inspection and report a final sample of 231 candidates. Similar to previous searches, they encountered the common issue of an extremely low true-positive rate of less than 1\%, which reflects the current state-of-the-art performance. 

Currently, the sample of strong lensing candidates consists of tens of thousands of systems \citep{Vernardos2024}, but with upcoming astronomical surveys such as Euclid and the Legacy Survey of Space and Time (LSST), this number is expected to increase to $10^5$ - $10^6$ \citep{barroso2024euclidearlyreleaseobservations, Collett_2015_forthcoming}. For the much rarer DSPLs the picture is similar: A handful are currently known (e.g., \cite{2022ApJS..259...27O, Shajib_2020_dspl, 2017ApJS..232...15D, Tanaka_2016, Gavazzi_2008}) but $\mathcal{O}(10^3)$ and $\mathcal{O}(10^2)$ DSPLs are expected to be found in Euclid \citep{Sharma_2023} and LSST \citep{desc_collab}, respectively. The main challenge for these surveys will be finding strong lenses and DSPLs amongst the billions of non-lenses that they will predominantly observe.

In the near future, astronomical surveys will generate a massive amount of data, making it essential to develop search methods with higher true-positive rates. Without new methods, an implausible amount of human time would be needed to visually inspect ML-selected strong lensing candidates. Our work has two main goals: (1) identifying strong gravitational lenses in the Dark Energy Survey (DES) and (2) developing a methodology that can better handle the demands of upcoming wide-field astronomical surveys. To achieve this, we choose the Vision Transformer (Subsection \ref{subsec:vit}) as our machine learning architecture and adopt Interactive Machine Learning (IML) to create a more comprehensive training sample (Subsection \ref{subsec:training_dataset}). Additionally, we conduct two independent strong lensing (SL) searches: one to find strong lenses in general and another one to target DSPLs. We implement an independent search for DSPL because DSPLs can exhibit very different morphology from single source lenses \citep{Collett_2015}, and ML searches designed for single-plane lenses often struggle to identify them. In one study comparing different lens searches, \cite{Metcalf_2019} found that all automated searches failed to identify a compound lens in their data. Only expert visual inspection correctly identified the DSPL.

This paper is structured as follows: In Section \ref{sec:methodology}, we summarize the methodology adopted in this work. Subsection \ref{subsec:DES dataset} describes the DES dataset, \ref{subsec:training_dataset} outlines how our training sample is constructed, Subsection \ref{subsec:vit} summarizes the architecture of the Vision Transformer, and Subsection \ref{subsec:training} covers the training process and the ML model’s performance on a testing sample and known catalogs of strong lenses. Section \ref{sec:results} presents the results from the ML search, along with two rounds of visual inspection: one involving citizen scientists on the platform Zooniverse, and the other with strong lensing experts. Finally, in Section \ref{sec:discussion}, we discuss the significance of these results and summarize our conclusions.

\section{Methodology}\label{sec:methodology}

Applying machine learning to detect strong gravitational lenses represents a standard vision classification task. The initial step in this process is constructing a training sample. Given the limited number of known lenses, simulations of strong lenses are employed as positive examples within training datasets. Conversely, negative examples are typically extracted from real data obtained from astronomical surveys. 

In this section, we outline our methodology and it is organized as follows: Subsection \ref{subsec:DES dataset} describes the DES cutout images used in this work for training and applying our ML algorithms. Subsection \ref{subsec:training_dataset} details the construction of our ML dataset for training, validating, and testing. Subsection \ref{subsec:vit} provides an overview of the Vision Transformer's architecture and Subsection \ref{subsec:training} delineates the training process and discusses the ML model's performance.

\subsection{The Dark Energy Survey Dataset}\label{subsec:DES dataset}

The Dark Energy Survey utilizes the Dark Energy Camera (DECam) \citep{Flaugher_2015_decan} on the 4-meter Victor M. Blanco Telescope at the Cerro Tololo Inter-American Observatory in Chile. DES mostly makes use of five photometric filters (\textit{g, r, i, z,} and \textit{Y}) that collectively cover the range \raisebox{0.5ex}{\texttildelow}398-1065 nm. The latest DES public data release (DR2) encompasses the total six years of DES operations (DES Y6), covering approximately 5,000 square degrees of the southern sky and cataloging roughly 691 million distinct astronomical objects \citep{Abbott_2021}. This data release has a median point-spread function Full Width at Half Maximum (FWHM) of \textit{g} = 1.11$\prime$, \textit{r} = 0.95'', \textit{i} = 0.88'', \textit{z} = 0.83'' and \textit{Y} = 0.90''. The median coadded catalog depth for a 1.95'' diameter aperture at signal-to-noise ratio = 10 is \textit{g} = 24.7, \textit{r} = 24.4, \textit{i} = 23.8, \textit{z} = 23.1, and \textit{Y} = 21.7 mag. 

In this work, we use \textit{g}-, \textit{r}-, and \textit{i}-band 45x45 pixel (\raisebox{0.5ex}{\texttildelow}12''x12'') cutouts of the Y6 coadded images to construct our training sample and apply our machine learning algorithms. The image sample for ML application is created by selecting objects from the Y6 catalogs that are likely to be galaxies, which are identified by applying a cut on the morphological star/galaxy classifier of \verb|EXT_COADD| $> 1$ \citep{Abbott_2021}. Additionally, we apply a magnitude cut, selecting objects with i-band AB magnitudes between 15 and 23.5. We highlight that in contrast to many previous SL searches, we do not apply color selection cuts. The resulting sample has an approximate size of 236 million cutout images. Section \ref{subsec:training} describes how the images are pre-processed for the ML model. 


\subsection{Constructing the Training Sample}\label{subsec:training_dataset}

Certain kinds of astronomical objects with features that may resemble strong gravitational lensing frequently appear as false positives in SL searches. Examples include spiral galaxies, ring galaxies, edge-on galaxies with close neighbors, etc. Some of these kinds of objects are more abundant in the universe than others, e.g. ring galaxies are much rarer than spiral galaxies. Consolidating all such potential false positives into a single negative training class risks under-representing less common types. This under-representation can hinder the ML model's ability to learn the distinctive features of each object type, reducing its effectiveness in distinguishing them from genuine strong lenses. To mitigate this issue, we designed our machine learning searches as multi-class image classification tasks, with specific training classes tailored to address these different types of common false positives.

The training sample was constructed implementing Interactive Machine Learning (IML), a form of human-in-the-loop ML methodology characterized by close interaction between the ML model and the developer. In IML, both entities exert comparable levels of control over the learning process. In this iterative approach, the performance of a trained model is evaluated, and the developer provides feedback by modifying either the training sample or the ML model’s features \citep{Amershi_Cakmak_Knox_Kulesza_2014}. By enabling humans to tailor input based on the model's current deficiencies, IML facilitates ML models to have predictions that better align with the developer's goals \citep{jiang2018recentresearchadvancesinteractive}.

\begin{figure*}[htbp]
 \centering
 \includegraphics[width=\textwidth]{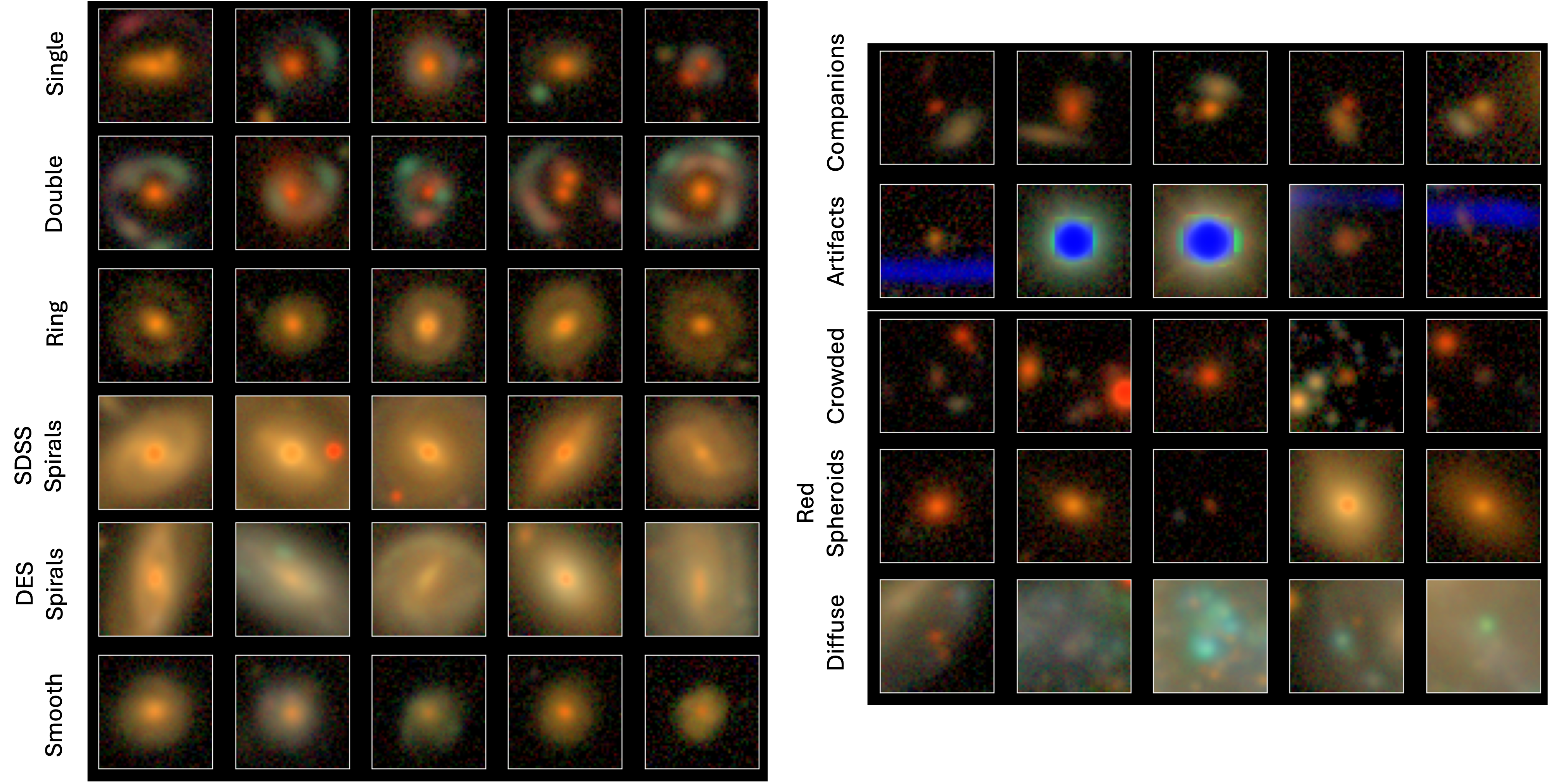}
\caption{Examples of the classification classes in our ML Dataset. This figure illustrates 5 example images from each of the 11 different training classes in our dataset. The examples demonstrate the distinct features within each class. Classes include `Single' (Single-plane simulations), `Double' (DSPL simulations), `Ring' (ring galaxies), `SDSS Spirals' and `DES Spirals' (spiral galaxies found in the respective survey), `Smooth' (diffuse, extended galaxies), `Companions' (central galaxies with adjacent companions), `Artifacts' (imaging anomalies), `Crowded' (densely populated regions), `Red Spheroids' (predominantly elliptical galaxies), and `Diffuse' (interstellar clouds of gas and dust). Each cutout measures \raisebox{0.5ex}{\texttildelow}12×12 arcseconds.}
\label{Fig:2ML_Dataset}
\end{figure*}

We implement IML by starting with two basic training classes for our primary search: positive (lens) and negative (not lens). When applying this initial trained model to real Y6 DES imaging data, an enormous quantity of images were incorrectly classified into the positive class exhibiting a true positive rate $\ll$ 1\%. A large fraction of these false positive cases were spiral galaxies and small and faint elliptical galaxies. As a consequence, we created a new training class for spiral galaxies and added faint elliptical galaxies to our initial negative class. We iterated this process several times and created a total of nine distinct training classes for the main search. These training classes were also included in the dataset for the DSPL search, supplemented by two additional classes: the target of this search (DSPLs) and an additional common type of false positive of this search. All training classes are described in detail in subsections \ref{sec:positive_classes} and \ref{sec:negative_classes}. Both training samples are similar in size, and in the DSPL search, single-plane strong lenses are part of a different training class than DSPLs. Figure \ref{Fig:2ML_Dataset} showcases example images for each training class. Additionally, Tables \ref{tab:training_sample_single} and \ref{tab:training_sample_double} show the entire composition of the ML dataset for the main and DSPL search, respectively. 

\begin{table}[htbp]
\centering
\title{test}
\caption{\\
Distribution of classes of the complete dataset for training, validating, and testing the main search ML model}
\label{tab:training_sample_single}
\begin{tabular}{l r}
\toprule
\textbf{Class}        & \textbf{Number of Images} \\
\midrule
Single                & 14,000 \\
Ring                  & 1,700 \\
SDSS Spirals          & 1,500 \\
DES Spirals           & 2,000 \\
Smooth                & 1,500 \\
Companions            & 1,000 \\
Artifacts             & 2,090 \\
Crowded               & 1,400 \\
Red Spheroids        & 15,000 \\
\midrule
\textbf{Total}        & \textbf{40,190} \\
\bottomrule
\end{tabular}
\tablecomments{\raggedright Subsections \ref{sec:positive_classes} and \ref{sec:negative_classes} describe the training classes.}
\end{table}

\begin{table}[htbp]
\centering
\caption{\\ Distribution of classes of the complete dataset for training, validating, and testing the DSPL search ML model}
\label{tab:training_sample_double}
\begin{tabular}{l r}
\toprule
\textbf{Class}        & \textbf{Number of Images} \\
\midrule
Double                & 8,000 \\
Single                & 8,000 \\
Ring                  & 1,600 \\
SDSS Spirals          & 1,500 \\
DES Spirals           & 2,000 \\
Smooth                & 1,250 \\
Companions            & 1,000 \\
Artifacts             & 2,090 \\
Crowded               & 1,400 \\
Red Spheroids        & 15,000 \\
Diffuse                 & 1,100 \\
\midrule
\textbf{Total}        & \textbf{42,940} \\
\bottomrule
\end{tabular}
\tablecomments{\raggedright Subsections \ref{sec:positive_classes} and \ref{sec:negative_classes} describe the training classes.}
\end{table}

While most examples in the negative training classes were sourced from previously published morphological catalogs of galaxies, approximately 8,000 examples were added through visual inspection of images that the ML model initially misclassified as positive. These examples were selected after manually reviewing tens of thousands of images, which required a significant time investment. However, by applying this ML model to the DES dataset, this work provides tens to hundreds of thousands of objects reliably categorized into each of our training classes, facilitating the creation of future training samples. Furthermore, recent efforts to refine galaxy morphology catalogs, such as \cite{walmsley2023galaxyzoodesidetailed}, will further reduce the need for extensive human involvement in the construction of training datasets.

\subsubsection{Simulated Positive Training Classes}\label{sec:positive_classes}

\begin{figure}[htbp]
 \centering
 \includegraphics[width=1\columnwidth]{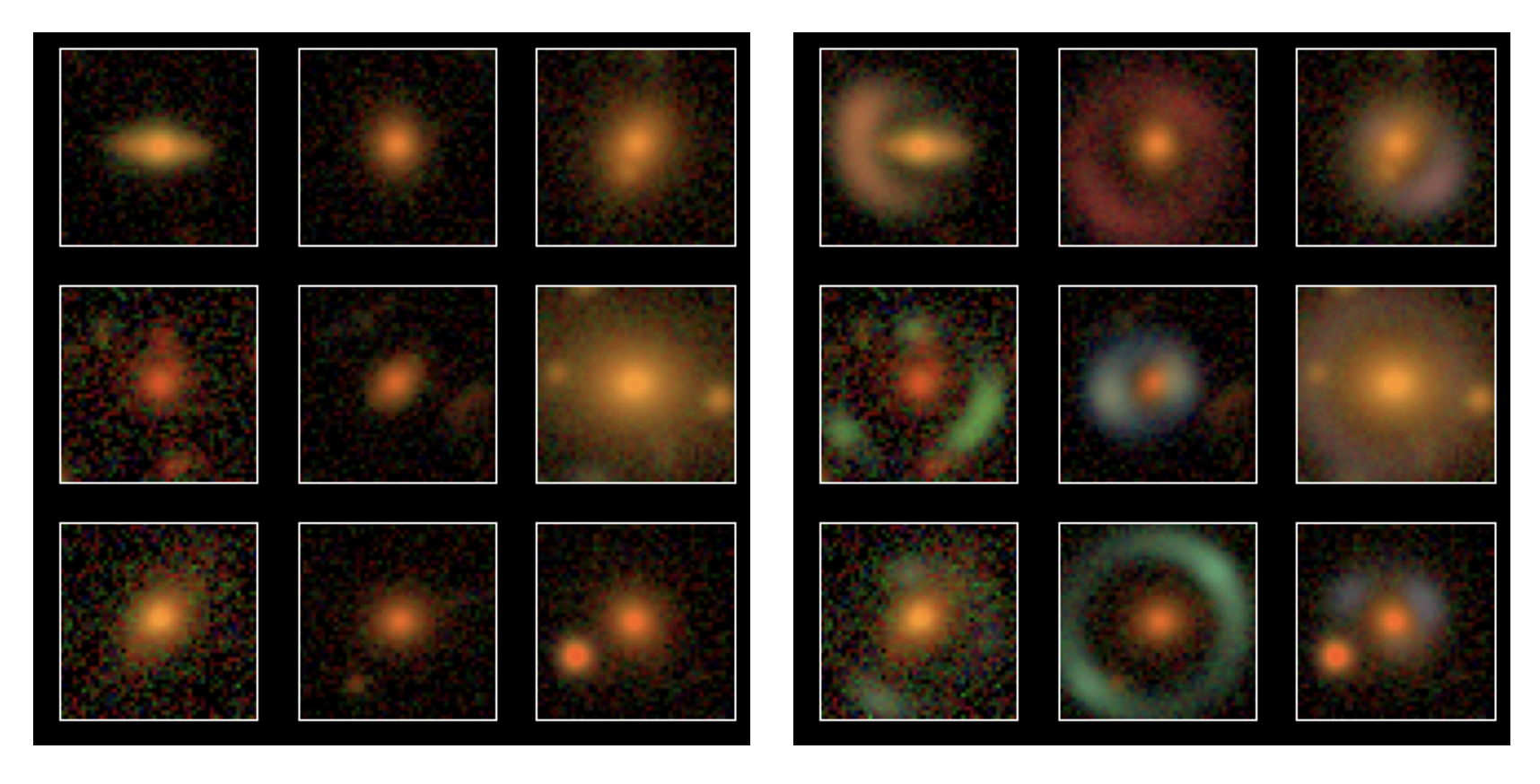}
\caption{Creation of simulations for the `Single' class. Left: Images of luminous red galaxies randomly selected from the DES redMaGiC catalog \citep{redmagic}. Right: The same images overlaid with simulated strongly lensed sources. Simulation properties are constrained to ensure obvious strong lensing features. Each cutout measures \raisebox{0.5ex}{\texttildelow}12×12 arcseconds.}
\label{Fig:3_simulations}
\end{figure}

\begin{figure}[htbp]
 \centering
 \includegraphics[width=1\columnwidth]{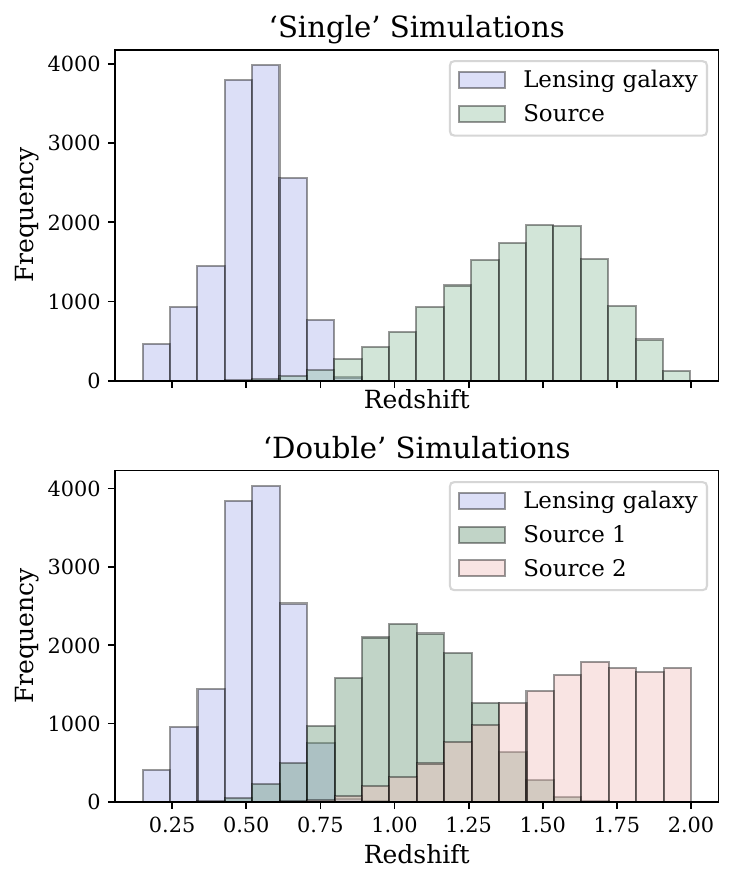}
\caption{Redshift distributions used for our strong gravitational lensing simulations. Top panel: `Single' class, ensuring the source redshift is at least 0.2 larger than the lensing galaxy's redshift. Bottom panel: `Double' class, ensuring the first source redshift is at least 0.2 larger than the lensing galaxy's redshift, and the second source redshift is at least 0.1 larger than the first source's redshift.}
\label{Fig:2_redshifts}
\end{figure}

`Single' is the positive class of our main search, while `Double' is the positive class for the DSPL search. Both classes are populated with simulations featuring real DES images of the lensing galaxies, overlaid pixel by pixel with images of simulated lensed sources. The lensing galaxies are selected randomly from a DES catalog created with the redMaGiC algorithm \citep{redmagic} that targets luminous red galaxies (LRGs), which tend to be massive elliptical galaxies. We consider only the \raisebox{0.5ex}{\texttildelow}50,000 galaxies with available spectroscopic redshifts, which are measured by external spectroscopic surveys as described in \cite{redmagic}. For this reason, the representation of deflectors with redshifts greater than 0.75 is limited (see Figure \ref{Fig:2_redshifts}), and this methodology is not expected to achieve a high completion rate for such systems. Additionally, we exclude galaxies with redshifts below 0.15 due to them being too large and bright for our cutout size. After choosing a lensing galaxy, we superpose simulated lensed sources into the image of the deflector galaxy, as shown in Figure \ref{Fig:3_simulations}. Finally, we randomly rotate and reflect the final image to augment this class with non-repeating simulations.

Through the adoption of IML, we observed a significant improvement in the performance of the ML model when utilizing simulations featuring clear strong lensing features. Consequently, we adopted simulation properties that produce bright and large lensed source images. Although the resulting simulations may not resemble realistic examples of strong lenses, this does not pose a problem in the context of a search as long as the selection function of the ML model is taken into account in population-level analyses of strong lenses. The use of IML likely tunes this methodology to preferentially identify strong gravitational lenses with large Einstein radii and bright lensed images.

The simulation of gravitationally lensed sources is generated using \href{https://github.com/lenstronomy/lenstronomy}{Lenstronomy} \citep{Birrer2021_lenstronomy, birrer2018lenstronomymultipurposegravitationallens}. The mass distribution of the lensing galaxy is assumed to be a Singular Isothermal Ellipsoid (SIE) profile, a simple mass profile capable of fitting most strong lenses. This mass profile requires as input parameters the velocity dispersion of the galaxy (attribute proportional to its mass), its position (assumed to be in the center of the cutout), and its ellipticity. The values of the velocity dispersion are drawn from a uniform distribution spanning the range 300-650 $km~s^{-1}$. For the ellipticity, we choose the orientation angle to be within 40 degrees of the galaxy's visible orientation angle, this is because the semi-major axis of the galaxy could be misaligned with respect to the orientation angle of the dark matter halo hosting it. The axis ratio of the ellipticity is allowed to vary uniformly between 0.001 and 1.

The redshift of the lensing galaxy is drawn from the sample of spectroscopic redshifts included in the redMaGiC catalog \citep{redmagic}. For the `Single' class, the redshifts of the deflector galaxy and the source follow the distributions shown in the top panel of Figure \ref{Fig:2_redshifts}, with the constraint that the source redshift is at least 0.2 higher than the lensing galaxy's redshift. 

The intrinsic light distribution of the source, prior to lensing, is modeled by a Sérsic light profile \citep{sersic1963influencia}. The eccentricity parameters, and surface brightness value at the half-light radius and the half-light radius for each photometric band are derived from measured properties of galaxies in the DES Y6 catalog with similar redshift values \citep{Abbott_2021}. The Sérsic index is randomly drawn from a uniform distribution between 0.3 and 4. In addition, we choose a random source position inside the region that produces multiple images with a total magnification higher than 3. The seeing of the simulated image in each photometric band is modeled by a Gaussian function with its FWHM determined using the DES Y6 survey condition maps for sky areas near the deflector galaxy. This Gaussian approximation is sufficient for our analysis, as it captures the dominant effects of seeing while maintaining consistency with the observational conditions.


\begin{figure}[htbp]
 \centering
 \includegraphics[width=1\columnwidth]{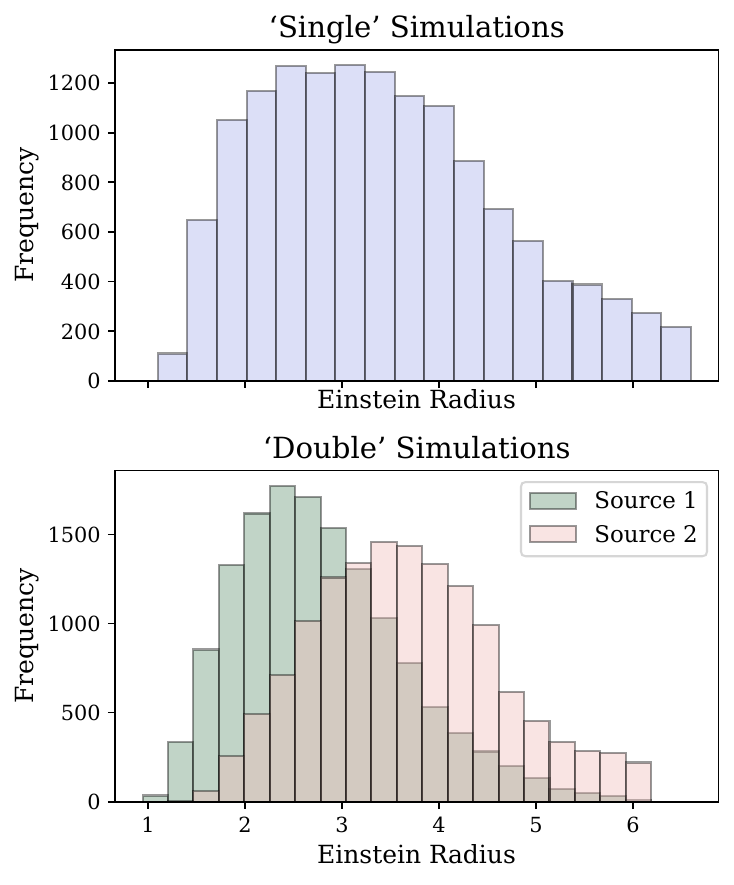}
\caption{Einstein radii distributions used for our strong gravitational lensing simulations. Top panel: 'Single' class. Bottom panel: 'Double' class.}
\label{Fig:2_einstein_radius}
\end{figure}

To make simulations with pronounced lensed arcs, we boost the brightness of the source by two magnitudes, similar to the adjustment made by \citet{2022AA...668A..73R} and \citet{Jacobs_b}. Additionally, we set constraints on simulation properties like the apparent magnitude of the lensed source, and source image position (correlated to the Einstein radius). The distribution of Einstein radii of these simulations is shown in the top panel of Figure \ref{Fig:2_einstein_radius}. Additionally, we define and constrain the values of two properties: `contrast' and `size comparison'. Contrast measures the brightness of the source relative to the lensing galaxy within the pixels occupied by the source. Size comparison estimates the potential overlap between the source images and the lensing galaxy as the ratio between the Einstein radius and the radius of a circle around the lensing galaxy's barycenter that contains half of its flux in the \textit{g} band.

In relation to the DSPL search, we are particularly interested in finding systems where the intermediate source galaxy is much less massive than the foreground lensing galaxy. This is because in this case lens modeling, the main systematic in strong lensing analyses, is much easier to determine. For this reason, we produce the `Double' simulations by generating two independent `Single' simulations with sources at different redshifts using the same lensing galaxy. Then, we overlay the image of the lensing galaxy with the simulations of both sources. Both `Single' simulations are required to pass the constraints described in the previous paragraph. The redshifts of the lensing galaxy and the sources follow the distributions shown in the bottom panel of Figure \ref{Fig:2_redshifts}, with the constraints that the first source redshift is at least 0.2 higher than the lensing galaxy's redshift and the second source redshift is at least 0.1 higher than the first source redshift. To ensure that the lensed features from each source are distinguishable and clear, we define and constrain the values of two new simulation properties. One of these constraints avoids simulations where the sources have colors that are too similar: we calculate the difference between the color indices $g - i$ and $g - r$ for the sources, and require that the sum of the absolute differences be at least 0.3 ($\left| (g - i)_2 - (g - i)_1 \right| + \left| (g - r)_2 - (g - r)_1 \right| \geq 0.3$). The other constraint discourages sources with similar Einstein radii by requiring a minimum absolute difference of 0.5'' in image separation between the values of both source simulations. The distribution of Einstein radii of these simulations is shown in the bottom panel of Figure \ref{Fig:2_einstein_radius}.


\subsubsection{Negative Training Classes}\label{sec:negative_classes}

As described in Section \ref{subsec:training_dataset}, our negative training sample was constructed employing an IML approach. We began with a basic negative training sample consisting solely of LRGs and gradually increased its complexity to enhance the ML model's performance on the DES dataset. While several training classes were populated using published morphological catalogs of galaxies, the majority were populated through the manual labeling of images that our ML models had incorrectly classified as either `Single' or `Double' with high probability.
 
Our main search uses the following negative training classes: Ring, Smooth, Companions, SDSS Spirals, DES Spirals, Crowded, Artifacts, and Red Spheroids. All of these classes have a degree of subjectivity and are best understood by referring again to Figure \ref{Fig:2ML_Dataset}. In summary, the `Ring' class consists of images showcasing ring galaxies, while `Smooth' denotes galaxies with a diffuse and extended appearance. `Companions' refer to images featuring a central galaxy with an adjacent edge-on or diffuse galaxy. `SDSS Spirals' and `DES Spirals' encompass mostly spiral galaxies originally identified within their respective astronomical surveys. These two classes were not combined because `DES Spirals' could not be filtered to include only face-on spiral galaxies, which are more challenging to distinguish from strong lenses, whereas `SDSS Spirals' was specifically constructed to include only face-on spirals. `Crowded' images capture densely populated regions. `Artifacts' represent images displaying imaging anomalies stemming from instrumental effects, processing errors, or transient phenomena. `Red Spheroids' comprises a diverse collection of images predominantly featuring elliptical galaxies. 

The DSPL search includes two additional negative training classes: Single (single plane lenses) and Diffuse. `Diffuse' images feature cloud-like, diffuse structures, often representing fields observed through the extended structure of a low-redshift galaxy. This class was added to this search because the initial DSPL ML model frequently misclassified such objects as 'Double'.

We populate some training classes with the SDSS morphological catalog presented in \cite{Dominguez_S_nchez_2018}, which comprises \raisebox{0.5ex}{\texttildelow}42,000 galaxies also identified in DES. From these \raisebox{0.5ex}{\texttildelow}42,000 galaxies, we select face-on spiral galaxies and classify them as either Ring (46 instances) or SDSS Spiral (317 instances). Both sets were augmented by applying rotations and reflections to increase their size to 100 and 1,100, respectively. Additionally, we select images with a high probability of displaying merging objects, augment this sample from 219 to 400, and add the resulting images to the SDSS Spiral class. The augmentation limits were chosen to increase sample size while avoiding over-representation and potential overfitting. Finally, we randomly select 2,000 galaxies likely to be round and add them to the Red Spheroid training class.

We populate more training classes with the morphological catalog presented by \cite{Cheng_2021}, which contains over 20 million galaxies detected within the DES footprint, along with a probability measure for each galaxy being elliptical rather than spiral. From this catalog, we select a sample of 2,000 galaxies likely to be spiral. Given the significant fraction of edge-on galaxies in this subset, we create the ‘DES Spirals’ class for them. This approach ensures that the class `SDSS Spirals’, allows the ML model to better distinguish the features of face-on spiral arms from lensing arclets. In addition, we add to the `Red Spheroids' class 2,500 highly probable elliptical galaxies ($p$ $>$ 0.9) and 2,000 moderately probable elliptical galaxies ($p$ $>$ 0.6), excluding previously selected objects.

The `Red Spheroids' class also includes 5,000 randomly selected redMaGiC galaxies \citep{redmagic} that we use as lensing galaxies in our simulations. Additionally, this class comprises 3,500 images that were classified by an early-trained ML model as having an extremely low probability of being strong lenses. 

Approximately 40\% of the `Artifact' class consists of images selected using columns from the Y6 GOLD table (Bechtol et al. in prep), which identify bright foreground objects or regions with missing data. 

The remaining images in the training set were selected through manual inspection and labeling of images incorrectly classified as `Single' or `Double' during the IML process. The distribution of labeled images across the different categories is as follows: 1,600 for `Ring', 1,500 for `Smooth', 1,000 for `Companions', 1,400 for `Crowded', 1,290 for `Artifacts', and 1,100 for `Diffuse'.

\subsection{The Vision Transformer}\label{subsec:vit}

\begin{figure*}[htbp]
 \centering
 \includegraphics[width=0.7\textwidth]{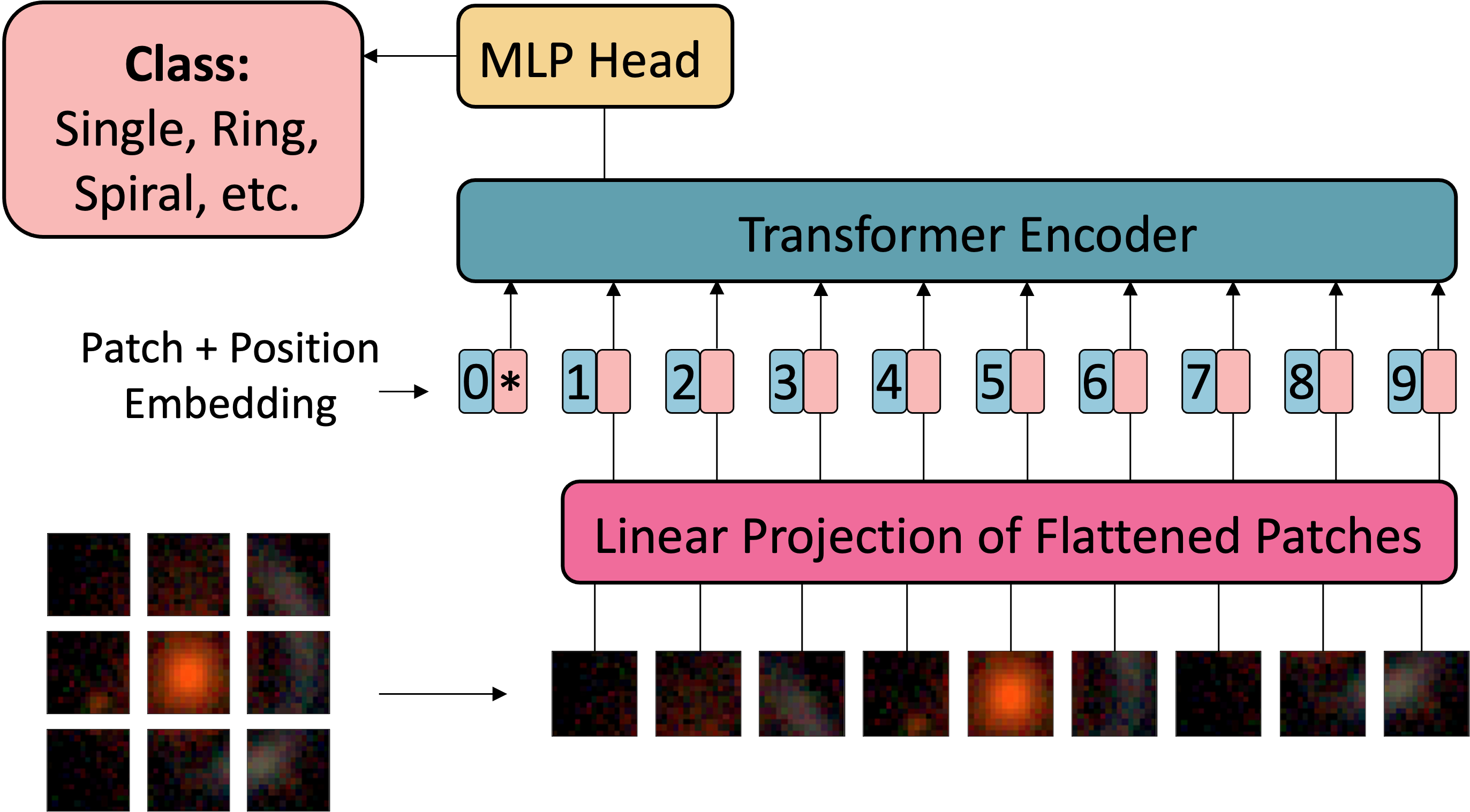}
\caption{Architecture of the Vision Transformer (ViT) \citep{vision_transformer}. The input image is divided into fixed-size image sections, which are linearly embedded and combined with position embeddings. These embeddings are then processed by a Transformer Encoder which applies multiple self-attention mechanisms. The output is finally fed into a classification head to predict the input image's class based on the training categories.}
\label{Fig:2_vision_transformer}
\end{figure*}

The machine learning model used for both searches is the Vision Transformer (ViT) \citep{vision_transformer}. We briefly describe the model's architecture here and refer the reader to the original paper for full details. Figure \ref{Fig:2_vision_transformer} provides a visual representation of the ViT architecture. First, the input image is divided into smaller, non-overlapping image sections called patches, with each patch projected onto a one-dimensional vector. Position embedding vectors are then added to these vectors to encode the position of each patch within the sequence. The resulting vectors, called `embedded patches', are then input into the Transformer Encoder, which consists of multiple stacked layers. The final element in the architecture is a Multilayer Perceptron (MLP) head that assigns the input image to one of the predefined categories for the classification task.

The Transformer Encoder was introduced in \cite{vaswani2023attentionneed} and is based on the self-attention mechanism, here represented by multi-head attention (MHA) blocks. The goal of the self-attention mechanism is to produce output vectors that are `context-aware' of the entire sequence. This is done by calculating them as a weighted sum of the input values using three matrices, whose components are learned during training. A multi-head attention block performs multiple attention functions in parallel and at the end, the output of each function is concatenated into a single vector. The final element in the Transformer Encoder is an MLP that acts as a classification head.

A key difference between transformers and conventional CNNs is their approach to image analysis. In CNNs, early layers focus on small, localized regions of an image, with each subsequent layer integrating information from progressively larger areas. In contrast, the Vision Transformer (ViT) processes the entire image from the initial transformer block. This allows the network to capture patterns across distant patches and identify global features early in the training process.


Since self-attention layers are global and lack the built-in assumptions about local patterns that CNNs have, they do not generalize well when trained on limited data. However, we can overcome this disadvantage by pre-training a ViT model on a large dataset and then fine-tuning it to our classification task, this is known as `transfer-learning' \citep{transfer_learning}. By using the learned features and weights of the pre-training process as a starting point, the amount of data and computational resources required for training is reduced and the performance of the model usually increases. When ViT is pre-trained on large datasets, it can perform the same or better than state-of-the-art CNN models (like BiT and Noisy Student) \citep{vision_transformer}.

In this work, we implemented the ViT model with the least number of parameters (ViT-Base, with 86 million parameters) and adopted a 16x16 pixel patch size. These choices reduce the computational resources required for training while maintaining high performance. Our ML model is pre-trained on ImageNet-21k \citep{imagenet}, which contains 14 million images and 21 thousand classes. Pre-training enables the model to learn general image features from a diverse dataset, providing a strong foundation that improves performance and reduces the data requirements for fine-tuning on our specific task. The model with its pre-trained weights is implemented using the Python library PyTorch Image Models ($Timm$) \citep{rw2019timm}. For the fine-tuning process, we add a single MLP layer at the end of the architecture to classify images according to the defined classes for our task.

\subsection{Training and Performance} \label{subsec:training}

The ML datasets used for both searches are divided into three subsets: 70\% for training, 15\% for validation, and 15\% for testing. Before being fed into the ML model, the images are resized to 224 x 224 pixels, as required by the pre-trained model available in the $Timm$ Python package \citep{rw2019timm}. The resizing is performed using the Resize function from the torchvision.transforms package, which employs bilinear interpolation. Additionally, the images are normalized to the mean values [0.485, 0.456, 0.406] and standard deviations [0.229, 0.224, 0.225] for the $g$-, $r$-, and $i$-bands, respectively. These values are used because the ML model was pre-trained on the ImageNet dataset, and these images were pre-processed using these specific mean and standard deviation values. During training, we minimize cross-entropy loss and stop training once the accuracy on the validation set fails to improve for more than three epochs. Training accuracy reflects the ML model's performance on the data it learned from, while validation accuracy provides an indication of how well the ML model generalizes to unseen data, helping to prevent overfitting. The ML model for the main search achieved training and validation accuracies of 95.5\% and 95.6\%, respectively, while the DSPL model achieved training and validation accuracies of 95.2\% and 93.9\%. We trained both ML models on the Open Science Pool \citep{osg_2015} platform, utilizing a GPU to accelerate the process from a few days to nearly four hours.

\subsection{Performance on the testing sample}

\begin{figure}[htbp]
 \centering
 \includegraphics[width=1\columnwidth]{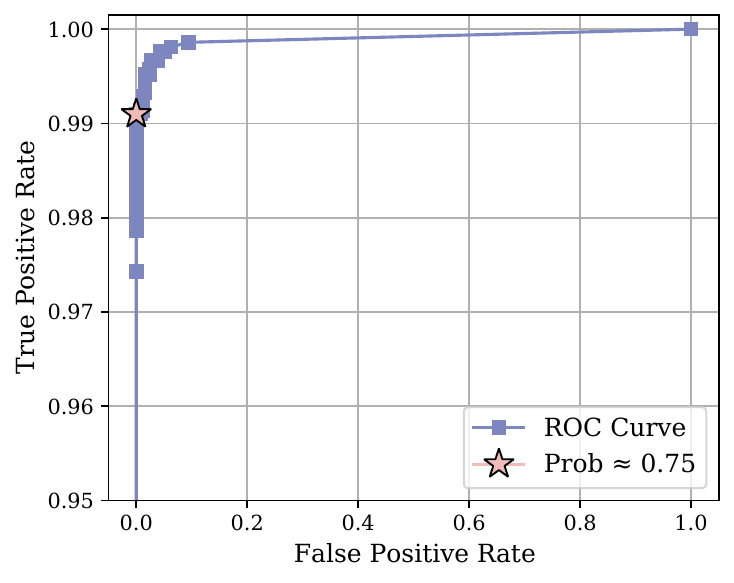}
\caption{Receiver Operating Characteristic (ROC) curve illustrating the performance of the main machine learning model on its testing sample. The pink star indicates the probability threshold chosen for application to the Dark Energy Survey dataset.}
\label{Fig:3_ROC_curve}
\end{figure}

We achieve classification accuracies of 95.2\% and 93.9\% on the testing samples for the main search and the DSPL search, respectively. Additionally, we assess the model's ability to distinguish between the positive class and any of the negative classes, yielding accuracies of 99.5\% for the main search and 99\% for the DSPL search. To further assess performance, we use a Receiver Operating Characteristic (ROC) curve, which illustrates an ML model's performance as a binary classifier as the discrimination threshold is adjusted. Figure \ref{Fig:3_ROC_curve} presents the exceptional ROC curve for the main model, where all negative classes are treated as a unified class.

Another performance metric for binary classifiers is the area under the ROC curve (AUC), which ranges from 0 (indicating a perfectly incorrect classifier) to 1 (indicating a perfect classifier), with a value of 0.5 representing random guessing. Both ML models demonstrate exceptional performance on their respective testing samples, achieving AUC values of \raisebox{0.5ex}{\texttildelow}1. While many previous ML strong lensing search studies have also reported near-perfect performance on their testing sets, the performance on real data differed significantly. Given that training, testing, and validation were conducted solely on simulated lenses and our negative sets are unambiguous, we should not anticipate achieving near-perfect performance on the actual DES data. This discrepancy highlights the importance of not overemphasizing model performance based solely on results from constructed testing sets.

\subsection{Performance on known SL candidate catalogs}

We use two catalogs of SL {\it candidates} identified in DES to evaluate the performance of our main ML model. The \cite{Jacobs_a, Jacobs_b} sample consists of 511 mostly galaxy-scale candidates discovered using CNN-based methods. The \cite{2022ApJS..259...27O} catalog contains 247 mainly cluster-scale candidates identified through non-ML algorithms. For this analysis, we visually inspect these candidates and exclude systems with lensing features outside our cutout images, resulting in 457 candidates from the Jacobs catalog and 140 from the O’Donnell catalog.

The ML model's output for an input image is a vector with components corresponding to the number of training classes, where each component, ranging from 0 to 1, represents the probability of the image belonging to that class. By classifying images into the class with the highest assigned probability, the model successfully recovers ~85.6\% of candidates from the Jacobs sample and 70\% from the O'Donnell sample. Given that neither the Jacobs nor the O'Donnell samples are completely pure, it is not surprising that many of their candidates are missed. Additionally, the lower recovery of the O'Donnell sample is probably due to group-scale lenses having more complex morphologies that are not represented in our training sample. 

\begin{figure*}[htbp]
 \centering
 \includegraphics[width=0.8\textwidth]{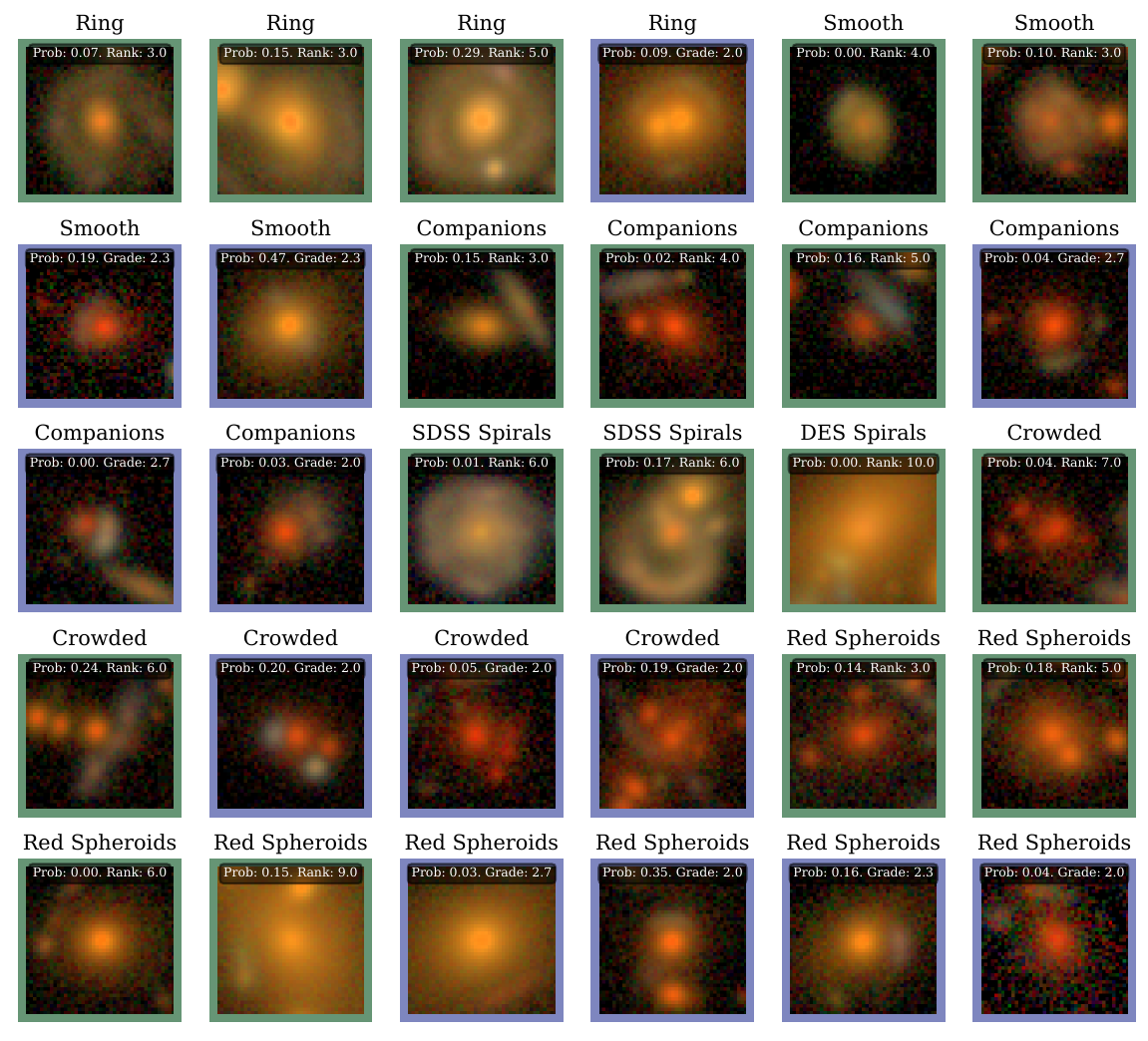}
\caption{Examples of strong lensing candidates \citep{2022ApJS..259...27O, Jacobs_a, Jacobs_b} not retrieved by the main ML model. Each image title indicates the assigned class, the model's probability of being `Single' (single-plane strong lens), and the expert-assigned `grade' or `rank' (0-3 for Jacobs, 0-10 for O'Donnell). Candidates from the O'Donnell catalog have a green frame, while those from the Jacobs catalog have a purple frame. Most unrecovered candidates lack definite strong lensing signs (with lower `grades' from \cite{Jacobs_a, Jacobs_b}) or group-scale strong lensing candidates. Each cutout measures \raisebox{0.5ex}{\texttildelow}12×12 arcseconds.}
\label{Fig:4_Known_lenses}
\end{figure*}

To provide a qualitative assessment of the ML model's performance, Figure \ref{Fig:4_Known_lenses} displays a collection of candidates that were not recovered. Each image title indicates the class to which it was classified, the model-assigned probability of being `Single', and an expert-assigned $grade$ (0-3) or $rank$ (0-10) from the Jacobs and O'Donnell samples respectively, reflecting the likelihood of being a strong lens. While \cite{Jacobs_a, Jacobs_b} required a $grade$ of 2 to qualify as a candidate, \cite{2022ApJS..259...27O}'s $rank$ cut was set at 3. Most of the candidates not recovered by the main ML model lack definite signs of strong lensing or seem to be group-scale lenses.

Without spectroscopic confirmation, it is not possible to determine whether our ML model fails to identify real lenses or correctly rejects false positives in the Jacobs and O'Donnell catalogs. Additionally, the Jacobs search was performed on a previous DES data release significantly shallower than DES Y6. Visual inspection suggests that many of the missed candidates are likely not true lenses, although some may be lenses that our classifier has overlooked. Nonetheless, most of the missed candidates remain ambiguous even after expert visual inspection.

\section{Results} \label{sec:results}

\begin{figure*}[htbp]
 \centering
 \includegraphics[width=0.9\textwidth]{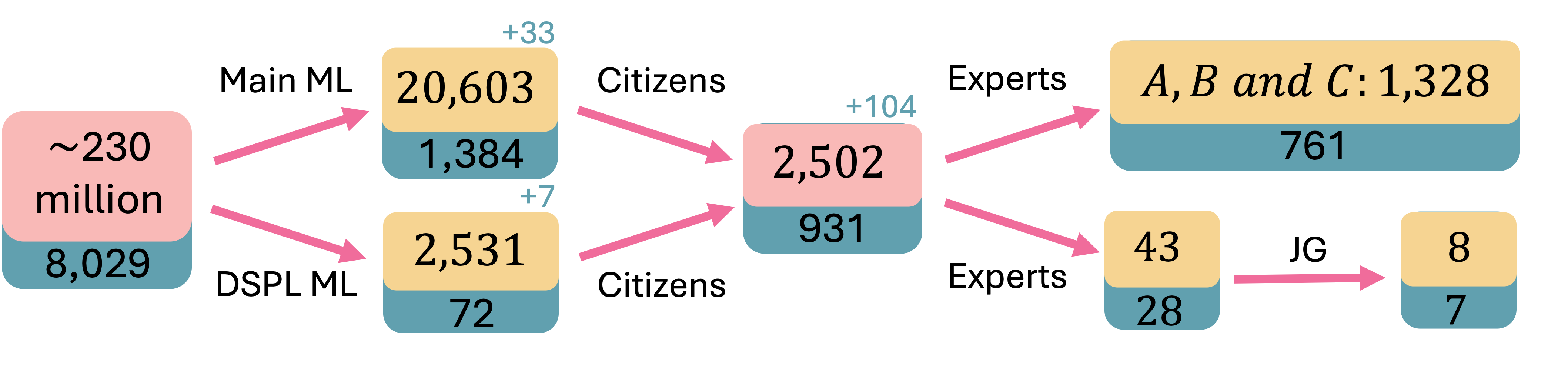}
\caption{Summary of our selection workflow, illustrating the reduction from 230 million images to 1,328 SL candidates and 8 DSPL candidates. Both ML models were applied to the DES cutout images, and those exceeding a chosen threshold were inspected by citizen scientists. The highest-scoring images were then reviewed by experts. The number at the bottom of each block represents the number of subjects reported in the SLED database as SL candidates including all reported scores and candidates identified in other surveys. We report 1,328 strong lensing candidates and 8 potential DSPL candidates.}
\label{Fig:5_summary_methodology}
\end{figure*}

We applied both ML models to the sample of \raisebox{0.5ex}{\texttildelow} 236 million DES cutout images and selected all images exceeding a threshold of 0.75 and 0.5 for the main and DSPL search, respectively. The output was then visually inspected by citizen scientists through a Space Warps\footnote{\url{www.spacewarps.org}} project on the Zooniverse\footnote{\url{www.zooniverse.org}} platform. Subsequent expert inspection was conducted on the resulting images. In Section \ref{subsec:ML_application}, we describe the application of ML to the DES dataset. Sections \ref{subsec:Visual_inspection_citizens} and \ref{subsec:results_visual_inspection} detail the setup and results of the citizen science inspection. Section \ref{subsec:expert_inspection} covers the expert visual inspection process and the classification of our final sample of candidates. Finally, in Section \ref{subsec:DSPL_results}, we present the best DSPL candidates identified. Figure \ref{Fig:5_summary_methodology} provides an overview of our selection workflow, illustrating the filtering of 230 million images to 1,328 SL candidates and 8 DSPL candidates. This figure includes the number of candidates that have been reported as strong lensing candidates in the SLED database \citep{Vernardos2024}, without doing any cuts on the reported expert score or on the astronomical survey where they were identified.

\subsection{Machine Learning}\label{subsec:ML_application}

We processed the \raisebox{0.5ex}{\texttildelow}230 million DES cutout images with both ML models on the FermiGrid\footnote{\url{https://www.fnal.gov/pub/science/computing/grid.html}}, and through parallelization, both applications took around three days to complete. For the main search sample, we selected images with a `Single' ML score of 0.75 or higher. For the DSPL search sample, we selected images with a `Double' ML score of 0.5 or higher. The `Single' probability threshold was chosen to obtain a reasonable number of candidates for visual inspection while still aiming to attenuate the presence of false positives. For the DSPL search, a much lower threshold was chosen as the focus in this case is to find true-positives. A significant fraction of the selected images contained the same SL candidates, albeit centered on different objects. We removed duplicated images and manually add 40 SL candidates visually identified during the IML process. This led us to a final sample of 20,636 `Single' candidates and 2,538 `Double' candidates, with 610 candidates in the intersection of both sets.

\subsection{Visual Inspection by Citizens}\label{subsec:Visual_inspection_citizens}

\begin{figure*}[htbp]
 \centering
 \includegraphics[width=0.9\textwidth]{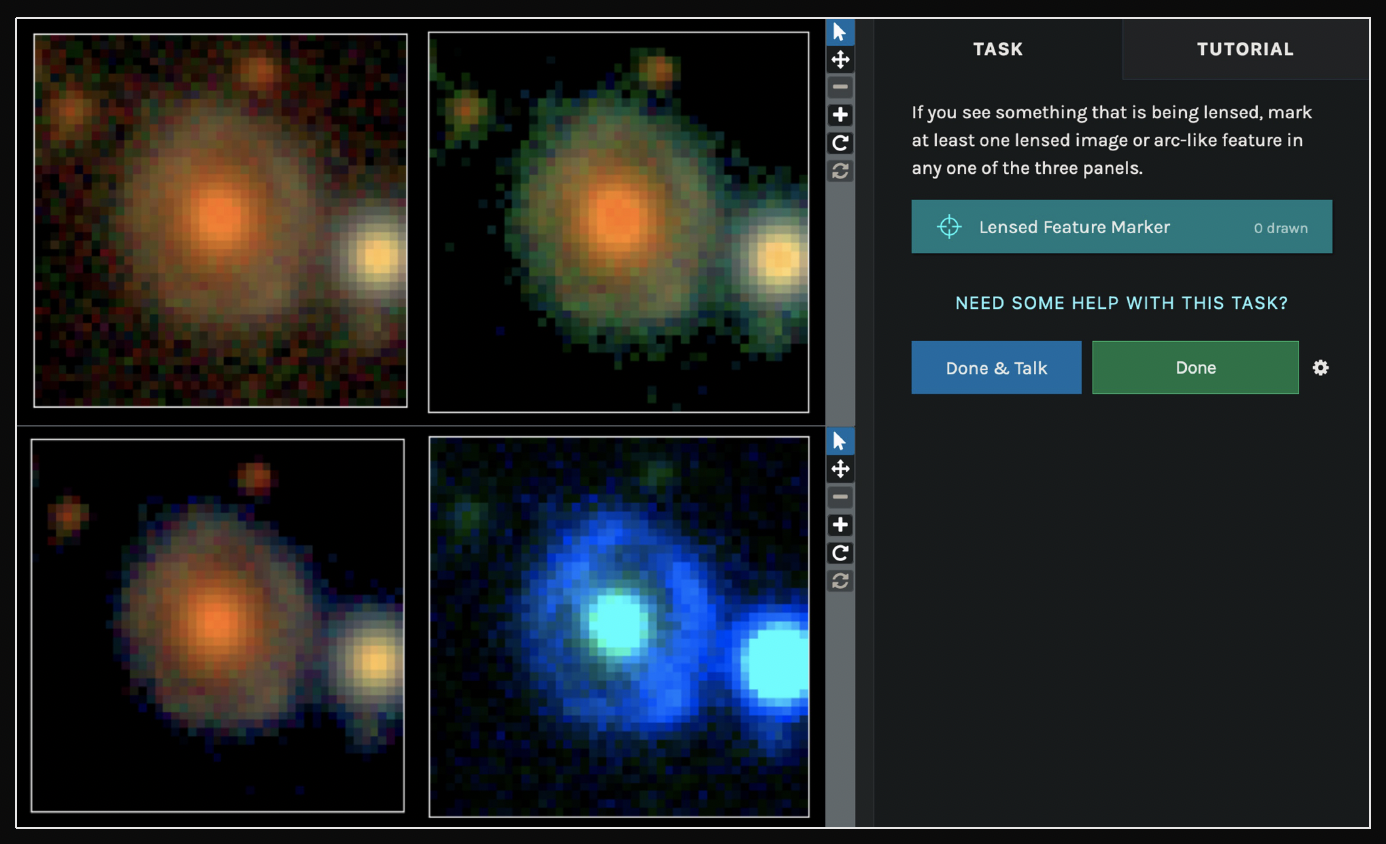}
\caption{Screenshot from the Space Warps citizen science project on Zooniverse. Participants were shown each candidate in four distinct PNG filters designed to highlight different image features and were asked to place at least one mark in the image if strong lensing features were recognized.}
\label{Fig:5_Zooniverse_screenshot}
\end{figure*}

The 22,564 candidates from both searches were visually inspected by citizen scientists as a Space Warps \citep{marshall_2016_sw, more_2016_sw} project, which ran from October 17th to November 3rd, 2023. During this period, 731 users participated with a median of 34 classifications. Each lens candidate was presented in four different `filters' or color balance PNG settings, which highlight different features in the images. Participants were asked to mark any strong lensing features they recognized in the image. The four PNG filters and the task instructions are shown in Figure \ref{Fig:5_Zooniverse_screenshot}.

In addition to our candidates, which we refer to as `test subjects', we included 2,451 `training subjects', comprised of 1,152 non-lenses, 444 known candidate lenses, and 855 simulated lenses. These known candidate lenses are extracted from the samples published in \citet{Jacobs_a, Jacobs_b} and \citet{2022ApJS..259...27O}, and did not include systems with strong lensing features outside our cutout images. We use the training subjects to provide active training to citizens during the classification task, providing live feedback indicating whether the classification was correct or incorrect after each classification was made. In addition, these classifications are used to calculate the users' success or skill in classifying lenses and non-lenses.

This method is summarised here, but readers are directed to \citet{marshall_2016_sw} for a complete description. The posterior probability $P_{k+1}(L)\equiv P(L|\{C_{U_0},...,C_{U_k}\})$ for a given training subject, having received $k+1$ classifications $\{C_{U_0},...,C_{U_k}\}$ from users $\{U_0,...,U_k\}$ is given by:
\begin{equation}
P_{k+1}(L)= \frac{P(C_{U_k}|L)\cdot P_k(L)}{P(C_{U_k}|L)\cdot P_k (L)+P(C_{U_k}|\hat L)\cdot P_k(\hat L)}
\end{equation}
where the classifications are either ‘Lens’ (`$L$’) or `Non-Lens’ (`$\hat L$’). The skill of each user is given by their responses to the training subjects they have seen, e.g, $P(C_{U_k}=$`$L$’$| L )\approx N_{\text{`L'}}/N_L$ is the ratio of the number of lens training subjects, $N_{\text{`L'}}$, that the user has correctly classified as a lens to the total number of lenses they have seen, $N_{L}$. In this manner, higher skill users can cause more significant changes in the posterior probability, while a user who mis-classified the training subjects 50\% of the time (i.e. random classification) would not change the subject score at all.

All subjects were assigned an initial score of $P_0(L)=5\times10^{-4}$ (the prior), based on the frequency of strong lensing in the galaxy population. These scores were processed via the Space Warps Analysis Pipeline (SWAP, \cite{Marshall2016}). If users reached a consensus on any test subject after at least 5 classifications such that its score was $p<1\times10^{-5}$ the subject was `retired' i.e. no longer shown on the platform. This is to maximise efficiency as the bulk of the task is to remove unlikely candidates. Test subjects with $p>1\times10^{-5}$ remained in the classification stream until 30 classifications had been made. Training subjects were not retired.

\subsection{Results of Visual Inspection}\label{subsec:results_visual_inspection}

\begin{figure}
 \centering
  \centering
  \includegraphics[width=0.4\textwidth]{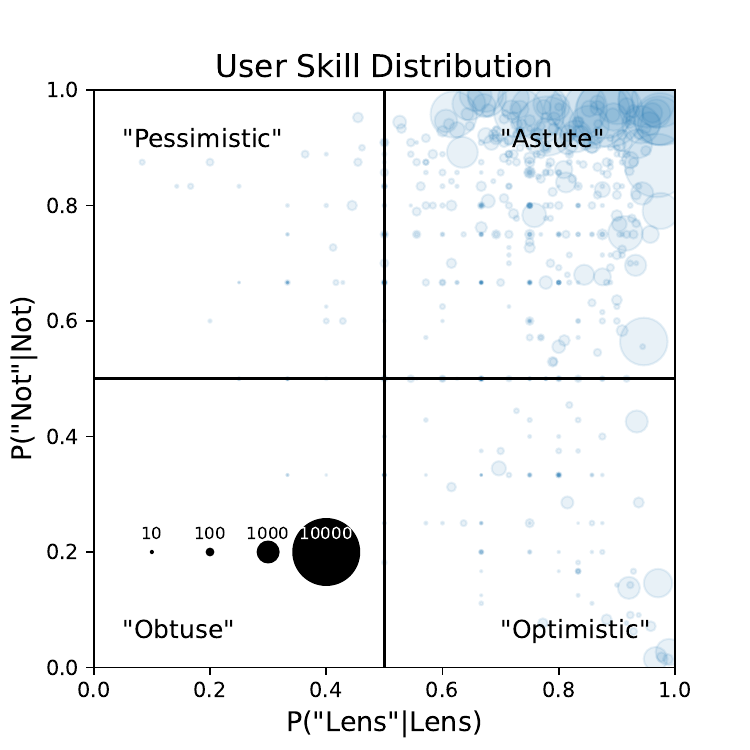}
\caption{A plot of the user skills, based on their performance on a training set. The user skill is defined as the fraction of correctly classified training subjects of each type, e.g the probability a user will classify a training subject as being a lens, given that it is indeed a lens, $P(`Lens$'$|Lens)$, and vice versa. The size of each datapoint is scaled by the number of classifications made by the user. The vast majority of users correctly identified most of the lenses and non-lenses they were shown.}
\label{Fig: User_skill_dist}
\end{figure}

The citizen scientists skill distribution is shown in Figure \ref{Fig: User_skill_dist}. The vast majority of classifiers were in the `Astute' quadrant of user skill, indicating that they could correctly identify the majority of training subjects, both lenses and non-lenses. The training subjects were accurately classified: most non-lenses received low scores, while most lenses (both simulated and real) received high scores. Furthermore, the majority of high-confidence real lens candidates also received a high score from the citizens.

Out of the 22,564 test subjects that were inspected by citizens, 2,502 received a score $p > 1\times10^{-5}$ and were subsequently inspected by SL experts. 


The Zooniverse platform allows users to engage in discussion boards about specific subjects, tag subjects with relevant labels (such as `lens', `possible', `double'), and save them to personal collections. During the process of classifying subjects, reviewing discussions, and examining various tags, JG saved `interesting' subjects into different collections. 104 of these subjects had received scores $p<1\times10^{-5}$ and were manually added to the pool of candidates to be inspected by SL experts.

\subsection{Expert Grading of Candidates}\label{subsec:expert_inspection}

Experts inspected the 2,502 subjects that received a score $p>1\times10^{-5}$ from citizens, along with 500 lower-scoring subjects randomly chosen for calibration purposes, and 104 candidates manually added by JG (as described in section \ref{subsec:results_visual_inspection}). The group of experts consisted of eight individuals: JAB, GC, PH, JG, MM, TL, KR, and SS. The inspection task was performed using the software Visapp \citep{more_2016_sw}. Each subject was displayed in the same four PNG settings we used in the visual inspection by citizens. Experts assigned a grade to each subject as follows:
\begin{itemize}
    \setlength\itemsep{0em}
    \item 0: Very unlikely ($<1\%$),
    \item 1: Probably not a lens ($2-50\%$),
    \item 2: Probable lens ($50-90\%$),
    \item 3: Certain lens ($>90\%$).
\end{itemize}
Additionally, we asked the experts to type the word `double’ in the Visapp comment box if they believed a subject could be a DSPL candidate.

\begin{figure}[htbp]
 \centering
 \includegraphics[width=1\columnwidth]{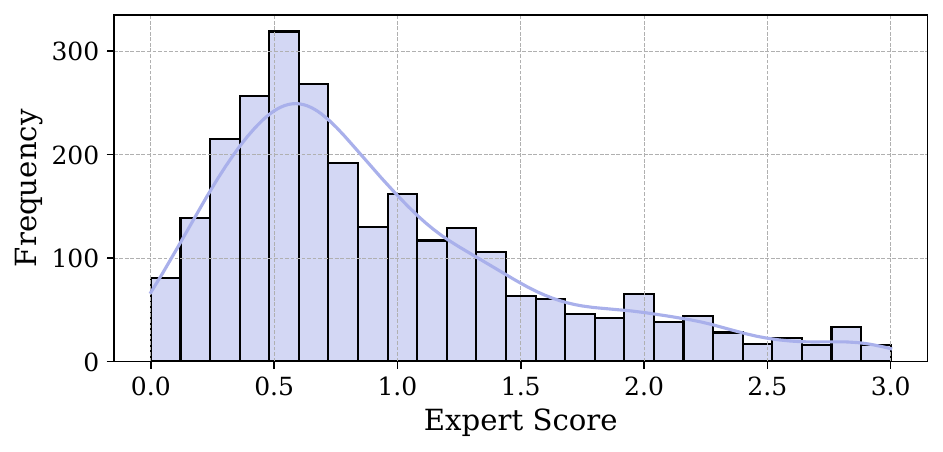}
\caption{Distribution of the calibrated expert scores for the 2,502 subjects with citizen scores exceeding $10^{-5}$, and the 104 subjects manually added by JG after reviewing Zooniverse discussion boards, tags, and personal collections.}
\label{Fig:5_Expert_distribution}
\end{figure}

We calibrated the scores from the experts and report this as the `expert score'. Figure \ref{Fig:5_Expert_distribution} displays the distribution of these values across the sample (not including the 500 random subjects). We use the following thresholds on the expert score for classifying our final 1,328 strong lensing candidates into categories A, B, and C:

\begin{itemize}
    \setlength\itemsep{0em}
    \item A: 2.25 $<$ Score (149 subjects, with 141 reported in the SLED database),
    \item B: 1.25 $<$ Score $<$ 2.25 (516 subjects, with 377 reported in the SLED database),
    \item C: 0.75 $<$ Score $<$ 1.25 (663 subjects, with 243 reported in the SLED database),
\end{itemize}

\begin{figure*}[htbp]
 \centering
 \includegraphics[width=0.9\textwidth]{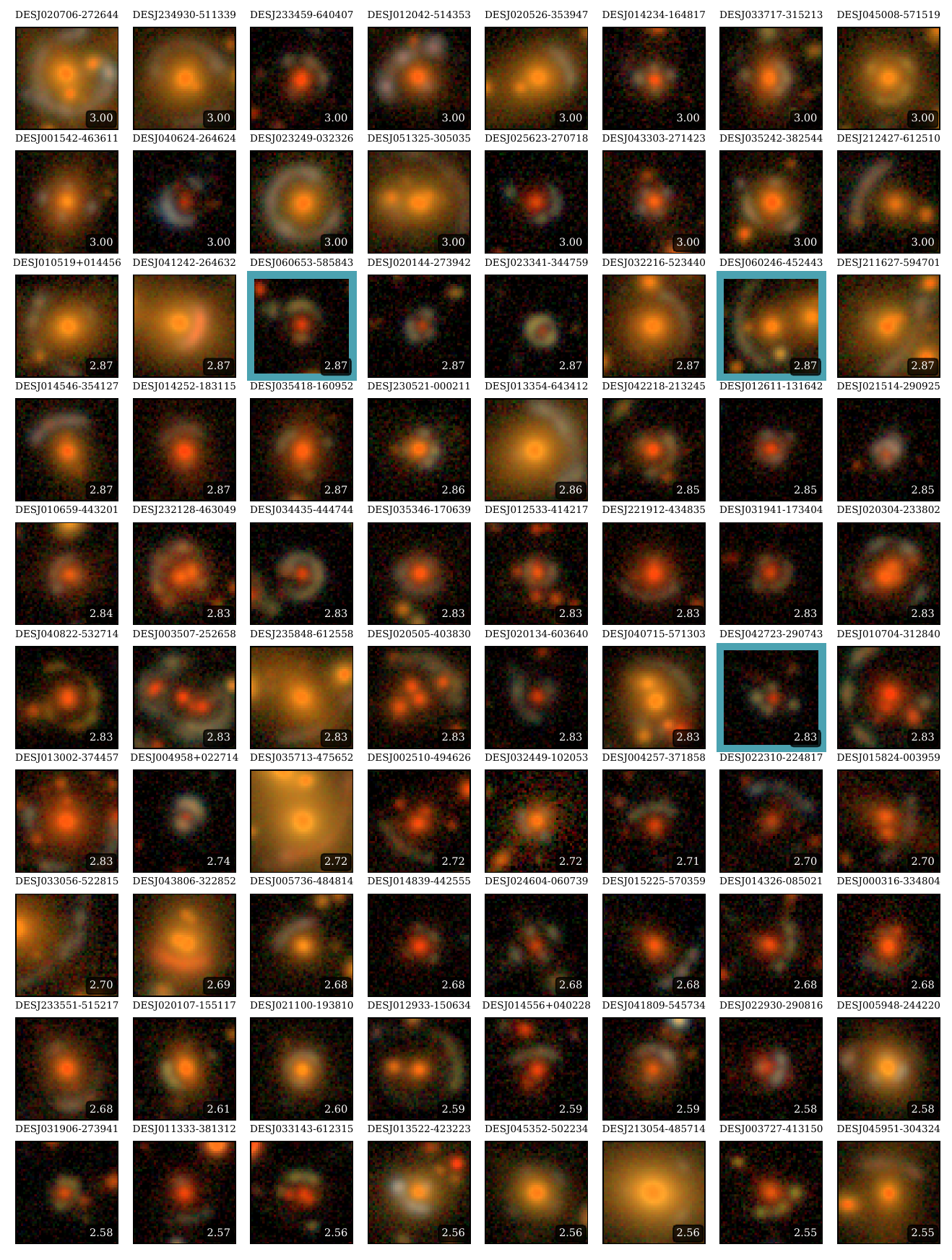}
\caption{Page 1 of the strong lensing candidates in our A category (expert score $>$ 2.25 out of 3). Eight of the candidates in this category have not been reported in the SLED database and are highlighted in blue in this figure. Each image shows the candidate’s coordinates and expert score, with each cutout measuring \raisebox{0.5ex}{\texttildelow}12×12 arcseconds.}
\label{Fig:5_A_candidates_1}
\end{figure*}

\begin{figure*}[htbp]
 \centering
 \includegraphics[width=0.9\textwidth]{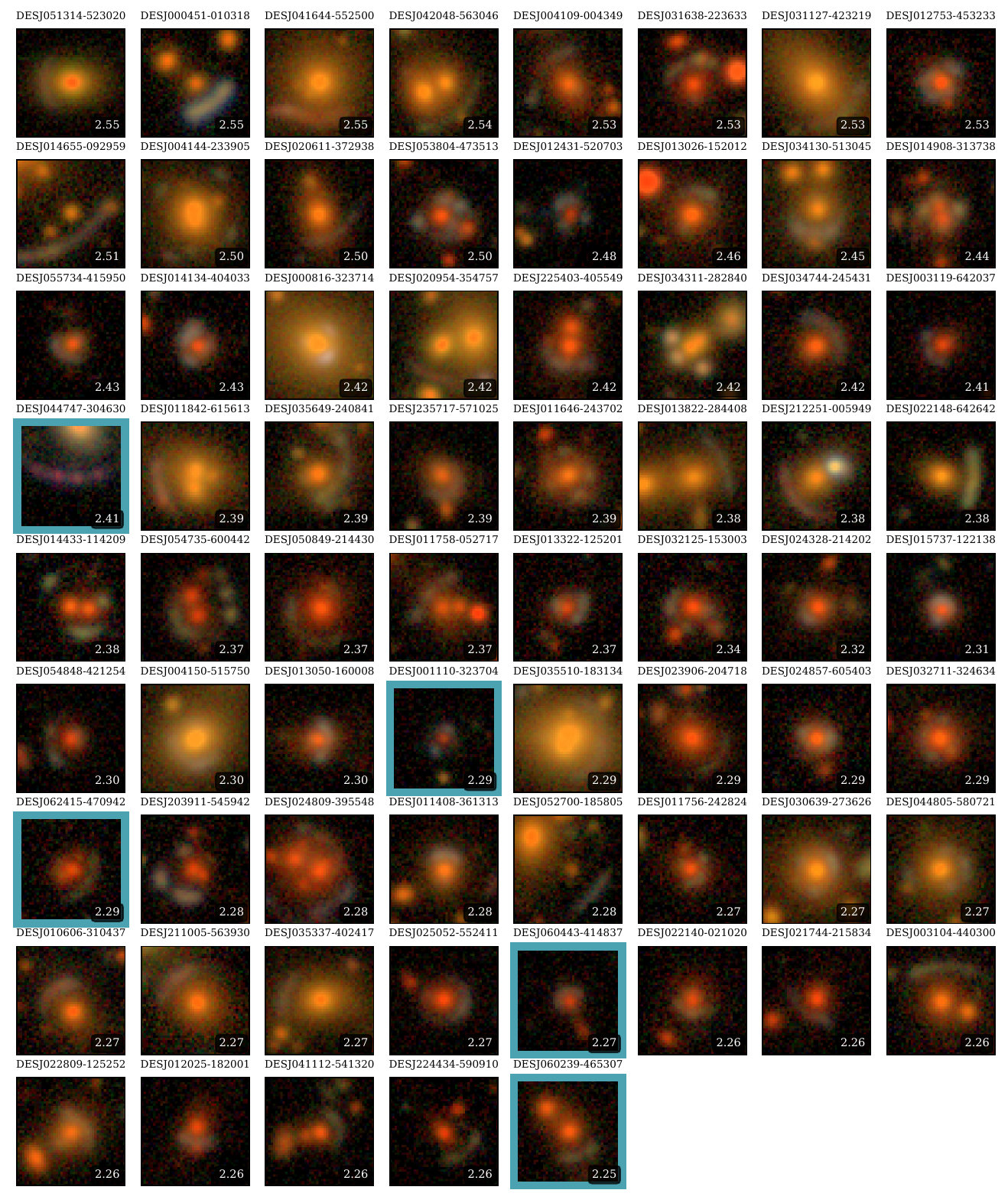}
\caption{Page 2 of the strong lensing candidates in our A category (expert score $>$ 2.25 out of 3). Eight of the candidates in this category have not been reported in the SLED database and are highlighted in blue in this figure. Each image shows the candidate’s coordinates and expert score, with each cutout measuring \raisebox{0.5ex}{\texttildelow}12×12 arcseconds.}
\label{Fig:5_A_candidates_2}
\end{figure*}

\begin{figure*}[htbp]
 \centering
 \includegraphics[width=0.92\textwidth]{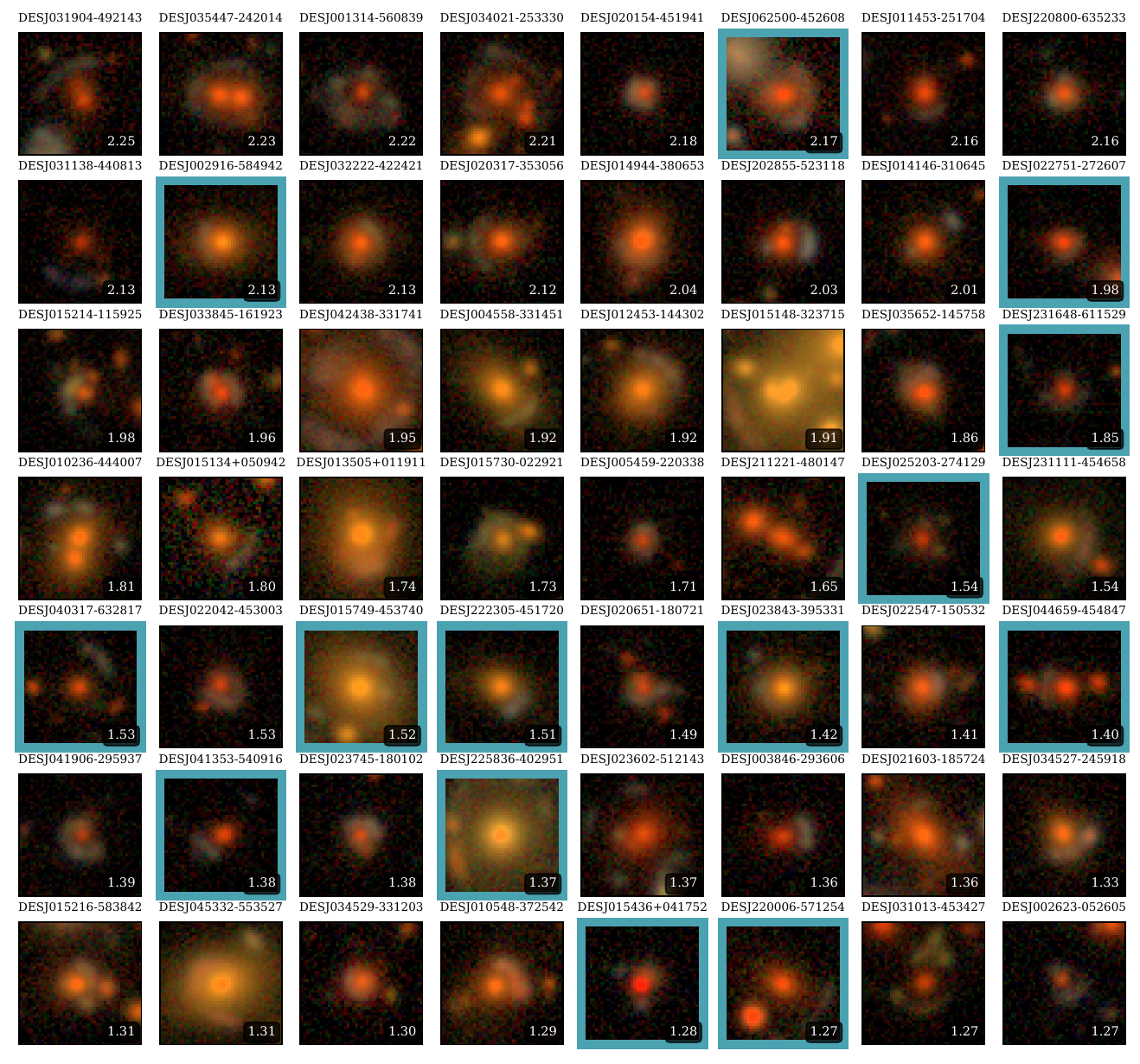}
\caption{Collage of 56 random strong lensing candidates in our B category (1.25 $<$ expert score $<$ 2.25, out of 3). Candidates not reported in the SLED database are highlighted in blue. This category contains 516 subjects. Each image shows the candidate’s coordinates and expert score, with each cutout measuring \raisebox{0.5ex}{\texttildelow}12×12 arcseconds.}
\label{Fig:5_B_candidates}
\end{figure*}

\begin{figure*}[htbp]
 \centering
 \includegraphics[width=0.92\textwidth]{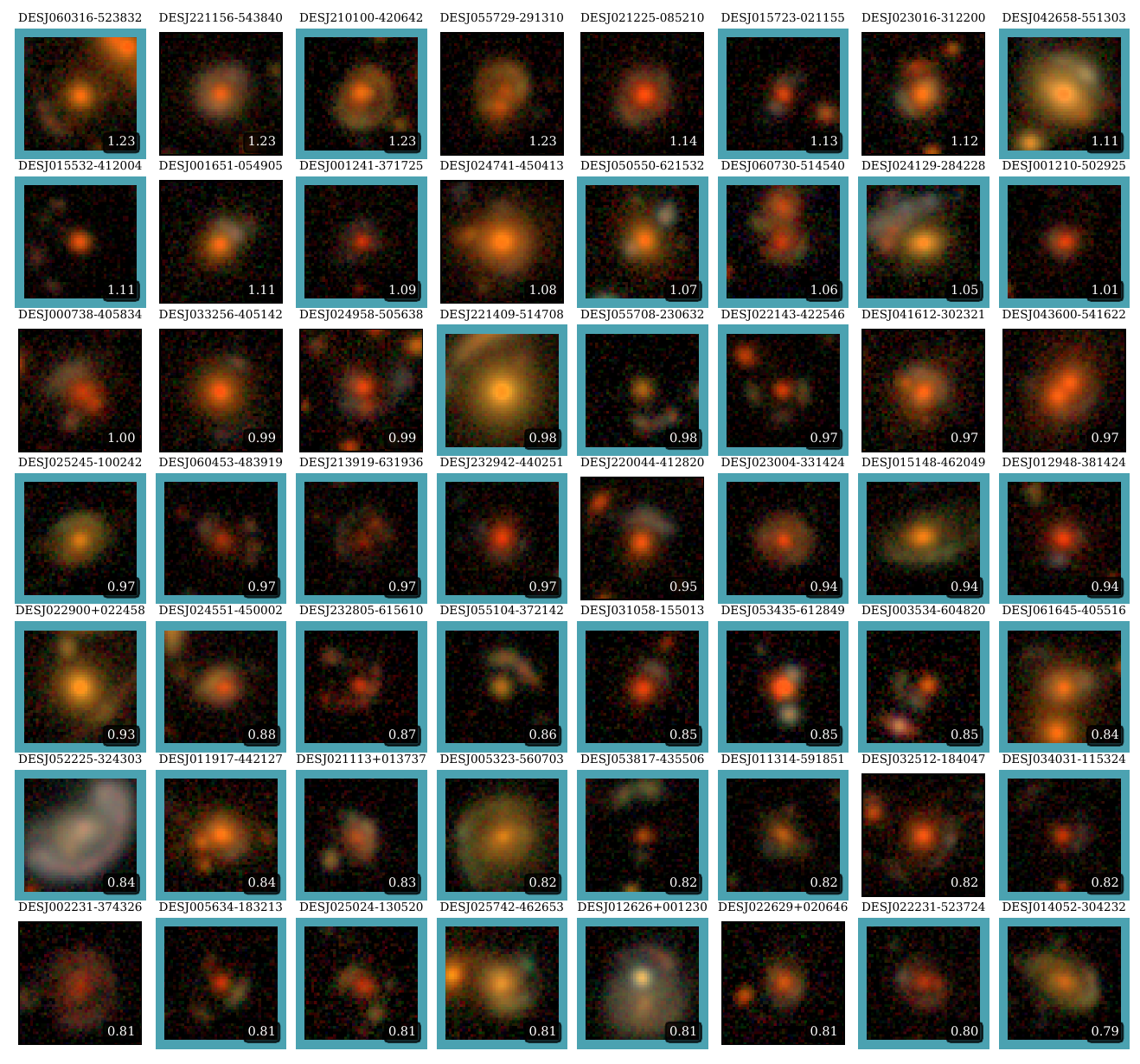}
\caption{Collage of 56 random strong lensing candidates in our C category (0.75 $<$ expert score $<$ 1.25, out of 3). Candidates not reported in the SLED database are highlighted in blue. This category contains 663 subjects. Each image shows the candidate’s coordinates and expert score, with each cutout measuring \raisebox{0.5ex}{\texttildelow}12×12 arcseconds.}
\label{Fig:5_C_candidates}
\end{figure*}

The A category contains primarily `definite' lenses, with all 149 subjects displayed in Figures \ref{Fig:5_A_candidates_1} and \ref{Fig:5_A_candidates_2}. Each subject in this category was scored a 2 or 3 by at least six of our seven experts, and at least two experts rated them a 3. Category B includes mostly `probable' lenses, with a subset of subjects shown in Figure \ref{Fig:5_B_candidates}. Most subjects in this category received a score of 2 or 3 from at least three experts. Category C comprises `maybe' lenses, which are less likely but could still be true positives. A subset of these subjects is shown in Figure \ref{Fig:5_C_candidates}. Each subject in this category was scored a 2 or 3 by at least one expert and exhibited a relatively high variation between the highest and lowest scores given by the experts.

\subsection{Double source plane lens candidates}\label{subsec:DSPL_results}

\begin{figure}[htbp]
 \centering
 \includegraphics[width=\columnwidth]{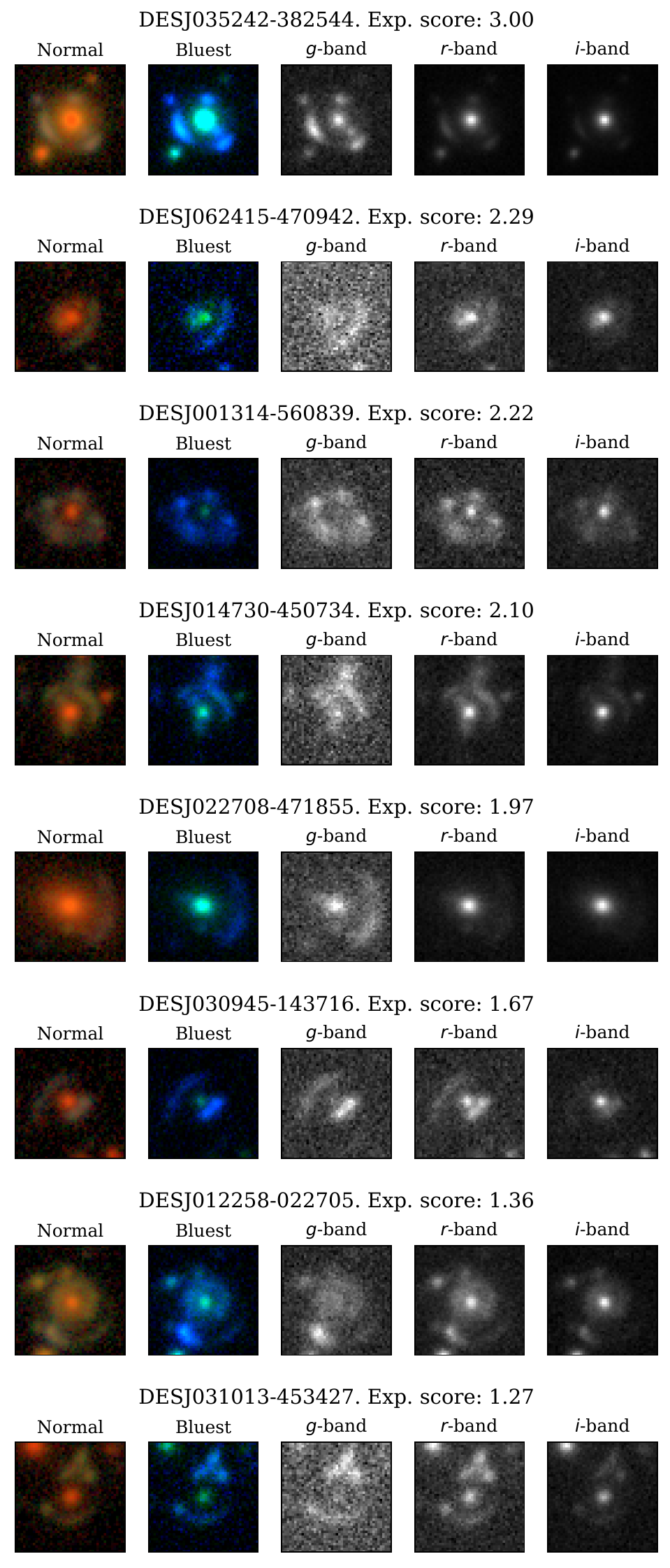}
\caption{The eight most promising galaxy-scale DSPL candidates from the 44 subjects identified by at least one expert as potential DSPLs. Each sub-figure’s title includes the candidate's coordinates and expert score. The first two images of each candidate are composites of the $g$-, $r$-, and $i$-band cutouts. Each cutout measures \raisebox{0.5ex}{\texttildelow}12×12 arcseconds.}
\label{Fig:5_best_DSPL}
\end{figure}

Out of the 2,502 subjects experts inspected, 44 received a `double' comment from at least one expert. This sample includes DESJ0610-5559, a multiple-source gravitational lens found previously in \cite{2017ApJS..232...15D} and \cite{2022ApJS..259...27O}. All clear multiple-source candidates in both \cite{2017ApJS..232...15D} and \cite{2022ApJS..259...27O} are cluster-scale systems, and given that our cutout size is not optimized for this scale, we only recovered one of them. JG reviewed the subset of 44 subjects to identify the most compelling galaxy-scale DSPL candidates, removing cluster-scale multiple-source lenses from consideration. Figure \ref{Fig:5_best_DSPL} presents the top 8 DSPL candidates with their respective expert scores across various PNG settings and bands. Notably, all these candidates were identified by the main ML model, with two (DESJ001314-560839 and DESJ012258-022705) also being selected by the DSPL model. Most of the candidates are probably at least single-plane lenses (in our A and B categories), which is also suggested by the observation that all candidates, except for DESJ012258-022705, have already been reported in the SLED database. DESJ035242-382544 is almost certainly a double-source strong lens, though spectroscopic data is needed to confirm whether the sources are on different planes. The remaining candidates exhibit characteristics indicative of potential DSPLs, but higher-resolution imaging and spectroscopic confirmation are required for definitive identification.

\section{Generation of an Ensemble Classifier}\label{subsec:ensemble_classifier}

Given classification scores were available for two different classifiers, the Vision Transformer and the Space Warps visual inspection, we tested whether an ensemble (combining the information from each individual classifier) would provide improved classification overall. With the arrival of wide-area surveys such as LSST, Euclid, and Roman wide-area surveys, the time-cost of expert grading of large numbers of high-scoring candidates will be significant but can be reduced by constructing the best possible performing classifier.

We generated an ensemble classifier via the `Isotonic Regression' method, the best-performing method studied in \citet{holloway2024}. We split the expert-graded sample equally into a calibration and test set. For the purpose of calibration, we defined a `true lens' as one receiving a grade $G\geq1.25$. We calibrated the scores of the Vision Transformer and citizen science project via Isotonic Regression using the calibration set to produce probabilities a given object was a lens; this process maps a given classifier score (e.g. a score of 0.9 from the Vision Transformer) to the average expert grade that subjects with that score receive (calculated via Isotonic Regression). We note due to our definition of a true lens in this case the `probabilities' are with respect to a system being an A or B grade; in the future a spectroscopic calibration set could be used to provide true probabilities of a lensing system. The probabilities determined from each classifier, $\{P_c\} = \{P_{SW}, P_{VT}\}$ are then combined via Bayes Theorem (see \citet{holloway2024} for a full derivation):
\begin{equation}
    P_{Ens}(L|\{P_c\}) = \frac{N_{NL}\cdot \prod_c P_{c}}{N_{NL}\cdot\prod_c {P_c} + N_{L}\cdot\prod_c (1-P_c)}
\end{equation}
where $N_L$ and $N_{\hat L}$ refer to the number of lenses (i.e. grade A+B) and non-lenses (including low confidence candidates in the C category) in the calibration set respectively.
\begin{figure*}
 \centering
  \centering
  \includegraphics[width=0.7\textwidth]{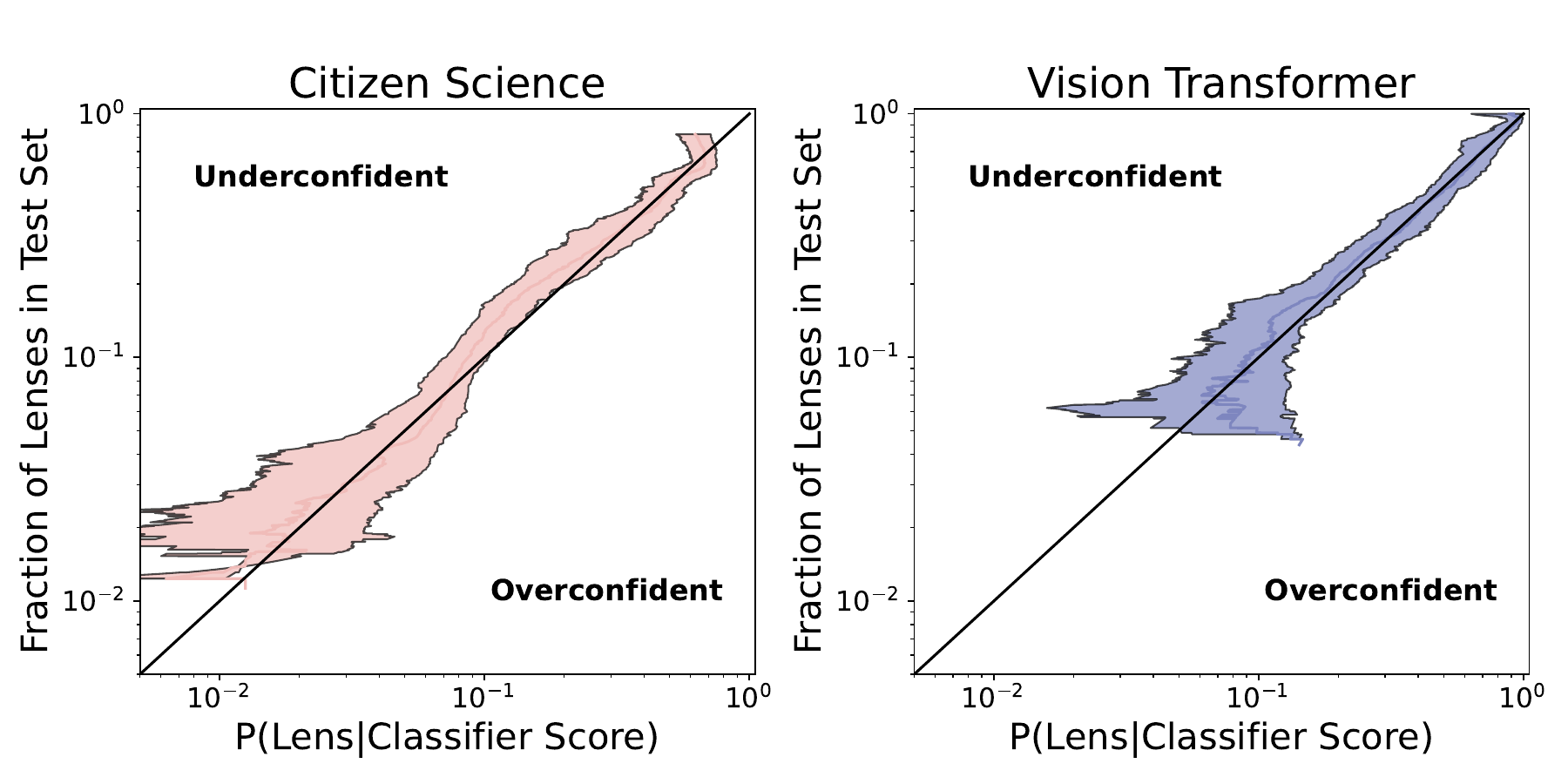}
\caption{Validation of the calibration curves generated via isotonic regression, performed on a separate test set. The x-axis shows the calibrated probability, while the y-axis shows the fraction of objects with that assigned probability which are indeed lenses. A perfect calibration curve would therefore lie along the $y=x$ line (solid black). The error bars have been calculated via bootstrapping. }
\label{Fig: Calibration_Validation}
\end{figure*}

\begin{figure*}
 \centering
  \centering
  \includegraphics[width=0.7\textwidth]{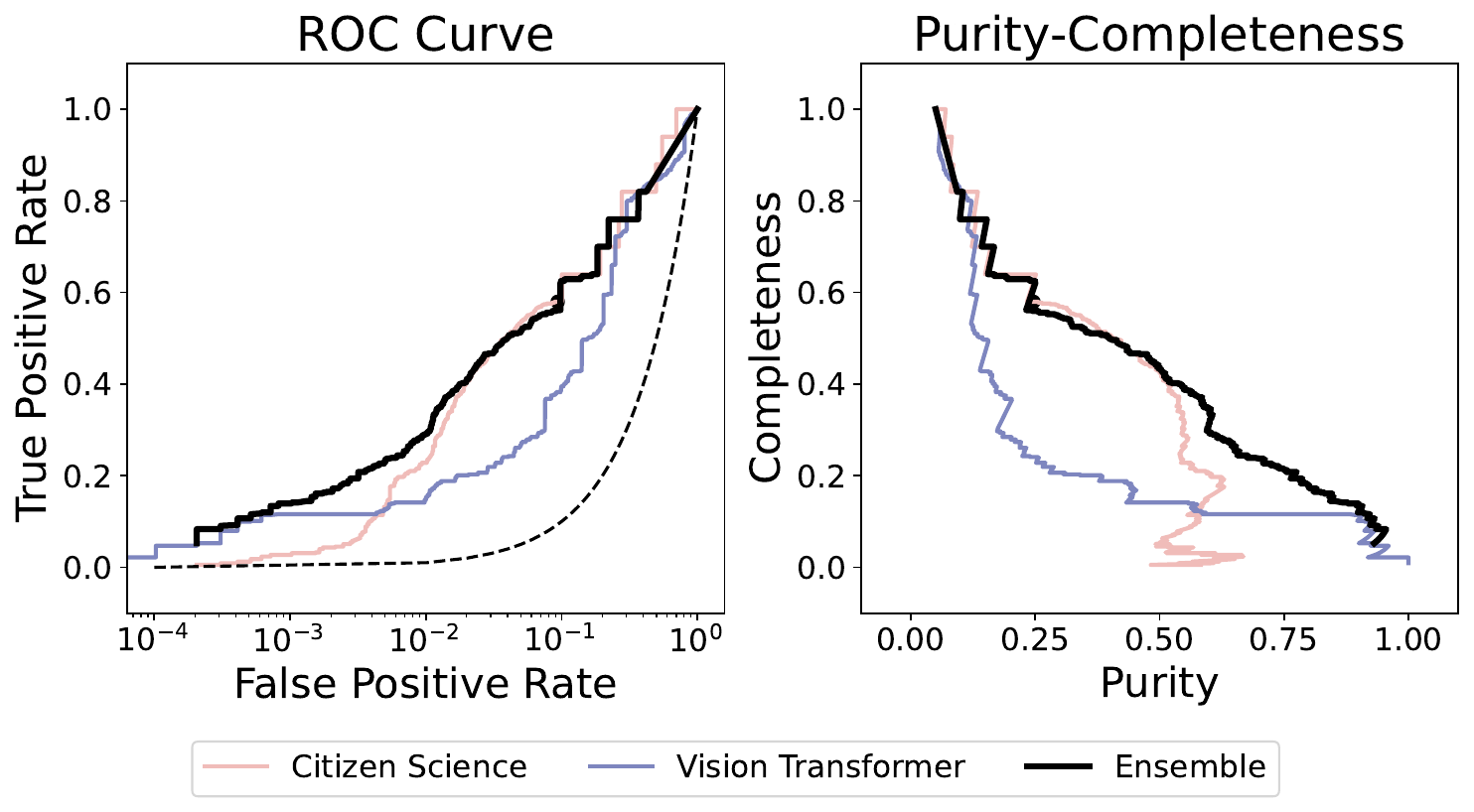}
\caption{Receiver operating characteristic (ROC) and purity-completeness curves for the individual and ensemble classifiers. The black point indicates the Space Warps threshold $p=1\times10^{-5}$, marking the point where expert grading moved from all subjects to a random subsample. The curve segments have been weighted to account for this difference. The dashed line indicates the ROC expected from a random classifier (note the left-hand x-axis is logarithmic).}
\label{Fig: PC_ROC_Curve}
\end{figure*}

Validation of this calibration was made on the distinct test set, which is shown in Figure \ref{Fig: Calibration_Validation}; a third `validation' set commonly used in machine learning was not required here as the Isotonic Regression method had no free parameters.  We then combined the calibrated probabilities for each subject from the vision transformer and citizen science search, assuming the two classifiers were independent \citep{holloway2024}. The ROC and purity-completeness curves of the individual and ensemble classifiers are shown in Figure \ref{Fig: PC_ROC_Curve}. We find the Citizen Science classifier typically provides greater completeness than the Vision Transformer, however the latter can provide a more complete sample when requiring high purity. We find our ensemble of just two classifiers can provide the best of both worlds, providing a performance at least as good as the best classifier for a given false positive rate (or purity). The ensemble classifier is dominated by the Citizen Science classifications at low purity ($\lesssim60\%$), while both classifiers play a significant role for higher purity thresholds. For future surveys this would reduce the number of systems required to be inspected; the ensemble classifier can provide lower false positive rates than the individual classifiers, and the calibration can be used to accurately predict the lens likelihood for each system in a dataset without them all being inspected.

\section{Discussion and Conclusions} \label{sec:discussion}

The main goal of this work was to develop an ML search methodology for strong gravitational lenses that can better handle the demands of the upcoming era of big data by producing samples of candidates with higher true-positive rates (TPR). We developed two ML models: one to find strong lenses in general and another to target DSPLs. We designed both searches as multi-class classification tasks, where we created specific training classes to address different types of potential false positives. We implemented a pre-trained Vision Transformer model as our ML architecture and adopted an Interactive Machine Learning approach to iteratively build and increase the complexity of our training sample.

Figure \ref{Fig:5_summary_methodology} summarizes our results. We applied both ML models to \raisebox{0.5ex}{\texttildelow}230 million DES cutout images, selecting the top 20,636 images from the main model and 2,538 from the DSPL model, with 610 images overlapping between the two samples. The selected images were visually inspected by citizen scientists, and the top 2,502 were further reviewed by experts, along with 104 images manually added by JG, as described in \ref{subsec:results_visual_inspection}. Out of these, 1,328 received an expert score higher than 0.75 out of 3, and 149 of these are grade-A candidates according to our expert scores. Figures \ref{Fig:5_A_candidates_1} and \ref{Fig:5_A_candidates_2} show all candidates in the A category. Figures \ref{Fig:5_B_candidates} and \ref{Fig:5_C_candidates} present a random collection of candidates in the B and C categories, respectively. The 8 most promising galaxy-scale DSPL candidates are shown in Figure \ref{Fig:5_best_DSPL}.

\begin{figure*}[htbp]
 \centering
 \includegraphics[width=\textwidth]{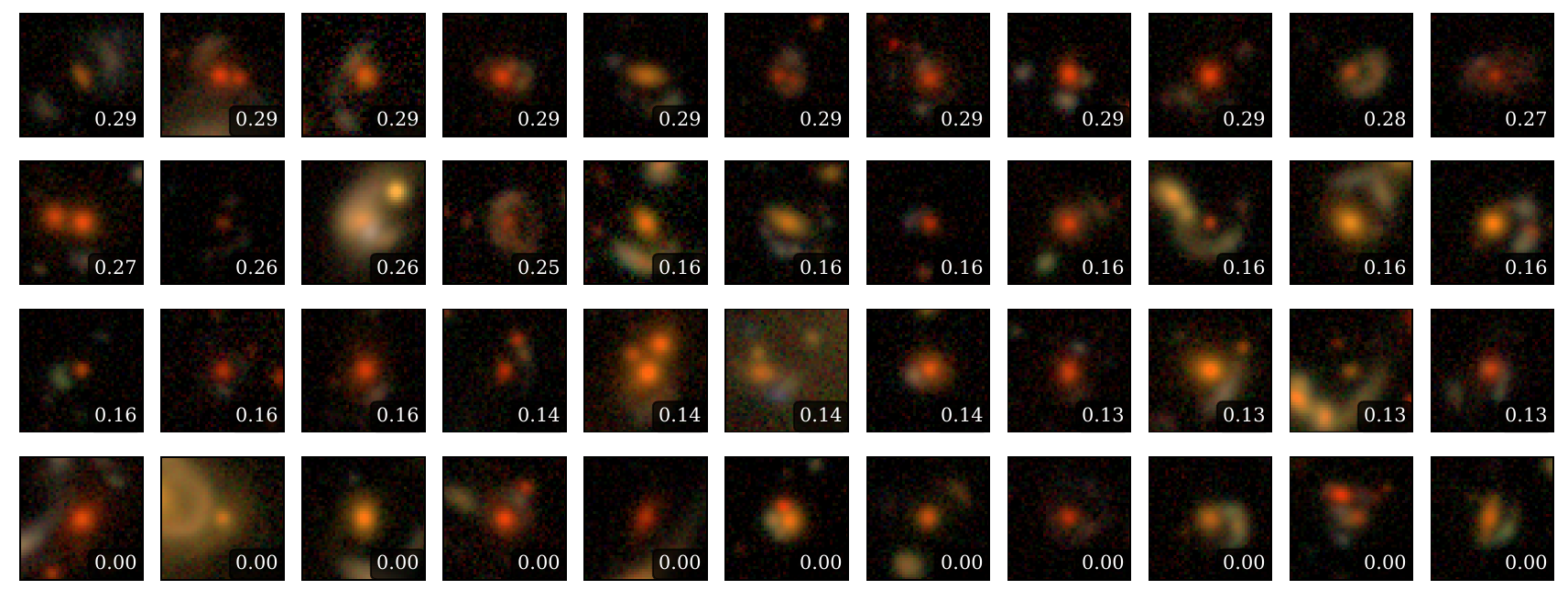}
\caption{Collection of 44 images that received very high scores from our main ML model ($\geq$ 0.998) but were rated low by the experts. These are challenging cases with features very similar to strong lensing, most of which appear to be galaxy mergers. The text box on the bottom right of each image is its expert score. Each cutout measures \raisebox{0.5ex}{\texttildelow}12×12 arcseconds.}
\label{Fig:6_False_positives}
\end{figure*}

Figure \ref{Fig:6_False_positives} presents a selection of 32 images that are scored highly by the main ML model ($\geq$ 0.998) and are rejected by the experts. The majority of these images exhibit features very similar to strong lensing. A significant fraction of them appear to depict merging galaxies and blue curved or elongated objects around a red galaxy, objects that are not well represented in our training sample. Future ML algorithms could be improved by including more of these types of systems in the training data or potentially by incorporating physics-based priors into the learning process.

To investigate the recovery rate of our methodology, we use four catalogs of strong lensing \textit{candidates} that have been identified in DES previously \citep{2022ApJS..259...27O, 2022AA...668A..73R, 2017ApJS..232...15D, Jacobs_a, Jacobs_b}. Results are summarized in Table \ref{tab:recovery_stats}. For the samples from Diehl and O'Donnell, we only consider systems with an Einstein radius lower than or equal to 6'' and a rank of 5 or higher out of 10, where 3 was the minimal rank to be considered a candidate in both works. Additionally, for the Jacobs sample, we do not consider systems with arc-lets outside our cutout images. For low-confidence candidates in Rojas, and all pre-selected candidates in Diehl, O'Donnell, and Jacobs, the main ML model alone recovers 52\%, 58\%, 72\% and 81\% of the samples, respectively. The low recovery rate for these Rojas candidates is expected as these are ambiguous systems without prominent lensing features or counter-images. The highest recovery rate is for the Jacobs sample, which is composed mostly of galaxy-scale SL candidates. In comparison, we noticed a lower recovery for Diehl and O'Donnell, which could be due to these samples containing group-scale candidates. Next, we consider only the high-confidence candidates from Rojas, and the candidates that pass the thresholds $grade$ $>$ 2.3 out of 3 and $rank$ $>$ 7 out of 10 from Jacobs, and Diehl and O'Donnell, respectively. We calculate a recovery of 76\%, 87\%, 88\%, and 90\% for the high-confidence candidates in O'Donnell, Diehl, Rojas and Jacobs, respectively. This proves that our ML model is successful in identifying galaxy-scale lenses with clear lensing features. 

The `(\%)' column of `Citizen Scientist (CS) Recovered' in Table \ref{tab:recovery_stats} considers only the systems that were recovered by the ML model and not the total pre-selected sample. Thus, this column offers an estimate of \raisebox{0.5ex}{\texttildelow}88.5\% for the completion rate of Citizen Scientists. When taking into account the selection by the main ViT model and the citizens' inspection, we calculate a recovery of 51\%, 62\%, 71\% and 81\% of the Diehl, O'Donnell, Jacobs, and Rojas high-confidence samples, respectively. 

We also investigated how many of our final high-confidence candidates were reported in the previous four catalogs of strong lensing candidates \citep{2022ApJS..259...27O, 2022AA...668A..73R, 2017ApJS..232...15D, Jacobs_a, Jacobs_b}. We calculate that the Diehl, O'Donnell, Jacobs and Rojas samples contain 10\%, 10\%, 39\% and 23\% of our candidates in categories A and B. Considering only the candidates in our A category, we calculate that the Diehl, O'Donnell, Jacobs, and Rojas samples contain 20\%, 23\%, 63\% and 40\% of these candidates, respectively. This highlights the relatively high completion achieved by our methodology.


\begin{table*}[htbp]
\caption{\\ Recovery statistics of our methodology for different strong lensing candidate catalogs identified in the Dark Energy Survey}
\centering
\label{tab:recovery_stats}
\begin{tabularx}{\textwidth}{l
                                >{\hfill\arraybackslash}X
                                >{\hfill\arraybackslash}X
                                >{\hfill\arraybackslash}X
                                >{\hfill\arraybackslash}X
                                >{\hfill\arraybackslash}X
                                >{\hfill\arraybackslash}X
                                r} 
\toprule
SL Catalog  & Total & \makebox[1.6cm][r]{Pre-selected} & \multicolumn{2}{c}{ML Recovered} & \multicolumn{2}{c}{CS Recovered} & \makebox[2.7cm][r]{ML\texttt{+}CS Recovered} \\ 
             &      &         & Count           & \text{(\%)}           & Count           & \text{(\%)}           & \text{(\%)}              \\ 
\midrule
\cite{2017ApJS..232...15D}  & 376  & 129          & 75              & \text{58.1}           & 66              & \text{88.0}           & \text{51.2}              \\
\cite{2022ApJS..259...27O}   & 252  & 94          & 68              & \text{72.3}           & 58              & \text{85.3}           & \text{61.7}              \\ 
\cite{Jacobs_a, Jacobs_b}   & 511    & 457          & 371             & \text{81.2}           & 329             & \text{88.7}           & \text{72.0}              \\ 
\cite{2022AA...668A..73R} (A) &90 & 90           & 79              & \text{87.8}           & 73              & \text{92.4}           & \text{81.1}              \\ 
\cite{2022AA...668A..73R} (B)  &315   & 315          & 164             & \text{52.1}           & 127             & \text{77.5}           & \text{40.3}              \\ 
\bottomrule
\end{tabularx}
\tablecomments{\raggedright Acronym definitions: ML = Machine Learning; CS = Citizen Science. The `\%' of `CS Recovered' is the percentage that citizens recovered from the sample selected by ML. `ML+CS' is the workflow of applying CS visual inspection to the output of ML. Rojas's A and B samples are high and low-confident strong lensing candidates, respectively.}
\end{table*}

Among our 228 candidates that received an expert score higher than 2, 71 were not reported by Jacobs or Rojas. Of these, 40 did not pass either of their color selection cuts. By not applying color cuts, we discovered $\sim$20\% more candidates but at the cost of analyzing $\sim$30 times more data. 

\begin{figure*}[htbp]
 \centering
  \includegraphics[width=0.7\textwidth]{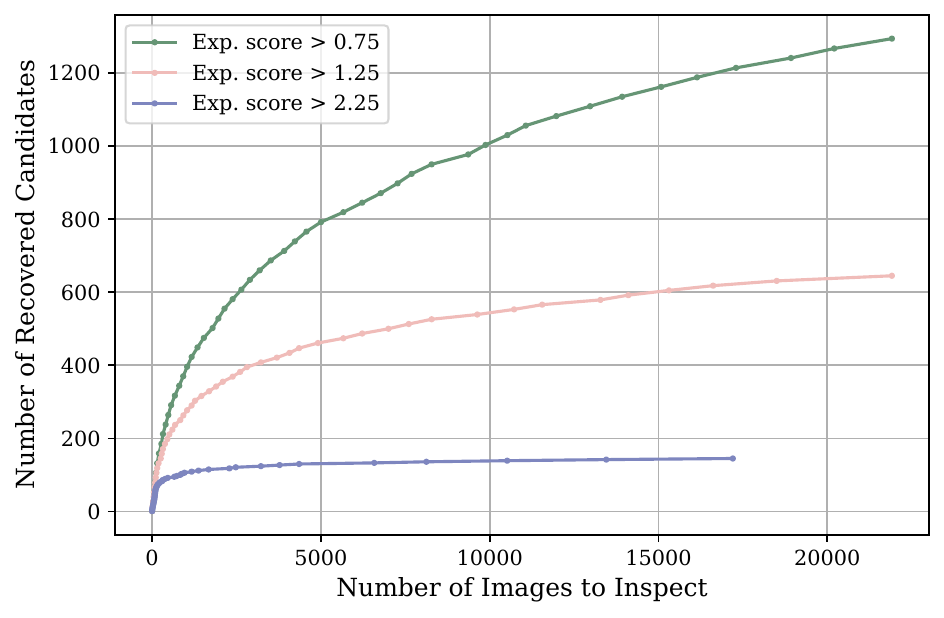}
\caption{Relationship between the number of images with ML scores exceeding a specified threshold (x-axis) and the number of good candidates among these images (y-axis). Candidates are separated by expert score thresholds. This figure illustrates the true-positive rate that can be achieved by our main ML model.}
\label{Fig:6_rec_vs_num}
\end{figure*}

The TPR that an ML model can achieve depends on the chosen score threshold, which determines the number of images to be inspected. Figure \ref{Fig:6_rec_vs_num} shows the number of images past a probability threshold that needs to be inspected (x-axis) to recover a certain number of candidates (y-axis). To recover 100 of the 149 subjects in our A category, we would need to inspect 830 images, representing a TPR of 12.1\%. A similar TPR of 12.8\% is reached to recover 400 of the 655 subjects with expert scores higher than 1.25. Discovering 600 of these would require inspection of 15,310 images (TPR of 3.9\%). These TPR estimates are much higher than previous searches, demonstrating the viability of our methodology for discovering tens of thousands of lenses in Euclid and LSST.

\section*{Data availability}

As part of this publication, we are making the following data products publicly available: 

\begin{itemize}
    \setlength\itemsep{0em}
    \item A dataset containing \raisebox{0.5ex}{\texttildelow}236 million scores from our main ML model \citep{gonzalez_2025_14588719}. This includes the coordinates of the systems and the complete probability vector for the different training classes. This dataset is highly valuable for constructing training samples in future strong lensing searches. The dataset is hosted on Zenodo and can be accessed \href{https://zenodo.org/records/14588719}{here}.
    \item Expert Scores: \raisebox{0.5ex}{\texttildelow}2,600 coordinates and expert scores of all subjects inspected by our experts.
    \item Ensemble Scores: \raisebox{0.5ex}{\texttildelow}20,600 coordinates and ensemble scores for all subjects inspected by citizen scientists.
\end{itemize}

\section*{Acknowledgements}

We extend our gratitude to the Zooniverse developers for their invaluable work in preparing and maintaining the platform, which was instrumental in facilitating this project. We also thank the citizen scientists for their vital contributions without which this project would be impossible. A full list of citizen scientists contributing to the classifications is given at the end of \href{https://www.zooniverse.org/projects/aprajita/space-warps-des-vision-transformer/about/team}{The Team} section of our Zooniverse project.


This research was done using services provided by the OSG Consortium \citep{osg_06, osg07, osg09, osg_2015}, which is supported by the National Science Foundation awards \#2030508 and \#1836650.

JG was supported by the U.S. Department of Energy through grant DE-SC0017647.

KB has been supported by the U.S. Department of Energy Early Career Program through grant DE-SC0022950.

This project has received funding from the European Research Council (ERC) under the European Union’s Horizon 2020 research and innovation programme (LensEra: grant agreement No 945536).

T.~E.~C. is funded by a Royal Society University Research Fellowship. 

SS has received funding from the European Union’s Horizon 2022 research and innovation programme under the Marie Skłodowska-Curie grant agreement No 101105167 — FASTIDIoUS.

J. A. A. B. acknowledges support from the Swiss National Science Foundation (SNSF).

Contribution statement: JG designed this project, implemented the Vision Transformer architecture, created and labeled the training sample, led the execution of the project and wrote this publication. JG, KB, SB, RM, ADW, JHO, and EZ contributed to producing the cutout images subsequently analyzed by the machine learning models. JG, PH, and AV prepared and ran the Space Warps visual inspection project on Zooniverse. PH and AV analyzed the results from the platform. PH also generated the ensemble classifier using machine learning and citizen science scores. TC contributed to the writing and organization of this publication. KB provided guidance and advice throughout the execution of the project. AV, PM, AM, Space Warps PIs, offered feedback on the design of the citizen science visual inspection project, the image display settings,  and the final ranking of candidates. JG, PH, AV, TL, CM, EMB and SB interacted with the citizens on ‘Talk’ responding to comments and helping citizens. JG, PH, JAB, GC, MM, TL, KR, and SS participated as experts in the second round of visual inspection. SB and TD served on the internal review committee for this publication, providing valuable feedback on the presentation of this work. SB and BN, co-conveners of the Strong Lensing group in DES, provided feedback on the execution of the project. The remaining authors contributed to this paper in various capacities, including the construction of DECam and other data collection instruments, data processing and calibration, developing broadly used methods, codes, and simulations, running pipelines and validation tests, and supporting the scientific analysis.


Funding for the DES Projects has been provided by the U.S. Department of Energy, the U.S. National Science Foundation, the Ministry of Science and Education of Spain, 
the Science and Technology Facilities Council of the United Kingdom, the Higher Education Funding Council for England, the National Center for Supercomputing 
Applications at the University of Illinois at Urbana-Champaign, the Kavli Institute of Cosmological Physics at the University of Chicago, 
the Center for Cosmology and Astro-Particle Physics at the Ohio State University,
the Mitchell Institute for Fundamental Physics and Astronomy at Texas A\&M University, Financiadora de Estudos e Projetos, 
Funda{\c c}{\~a}o Carlos Chagas Filho de Amparo {\`a} Pesquisa do Estado do Rio de Janeiro, Conselho Nacional de Desenvolvimento Cient{\'i}fico e Tecnol{\'o}gico and 
the Minist{\'e}rio da Ci{\^e}ncia, Tecnologia e Inova{\c c}{\~a}o, the Deutsche Forschungsgemeinschaft and the Collaborating Institutions in the Dark Energy Survey. 

The Collaborating Institutions are Argonne National Laboratory, the University of California at Santa Cruz, the University of Cambridge, Centro de Investigaciones Energ{\'e}ticas, 
Medioambientales y Tecnol{\'o}gicas-Madrid, the University of Chicago, University College London, the DES-Brazil Consortium, the University of Edinburgh, 
the Eidgen{\"o}ssische Technische Hochschule (ETH) Z{\"u}rich, 
Fermi National Accelerator Laboratory, the University of Illinois at Urbana-Champaign, the Institut de Ci{\`e}ncies de l'Espai (IEEC/CSIC), 
the Institut de F{\'i}sica d'Altes Energies, Lawrence Berkeley National Laboratory, the Ludwig-Maximilians Universit{\"a}t M{\"u}nchen and the associated Excellence Cluster Universe, 
the University of Michigan, NSF NOIRLab, the University of Nottingham, The Ohio State University, the University of Pennsylvania, the University of Portsmouth, 
SLAC National Accelerator Laboratory, Stanford University, the University of Sussex, Texas A\&M University, and the OzDES Membership Consortium.

Based in part on observations at NSF Cerro Tololo Inter-American Observatory at NSF NOIRLab (NOIRLab Prop. ID 2012B-0001; PI: J. Frieman), which is managed by the Association of Universities for Research in Astronomy (AURA) under a cooperative agreement with the National Science Foundation.

The DES data management system is supported by the National Science Foundation under Grant Numbers AST-1138766 and AST-1536171.

The DES participants from Spanish institutions are partially supported by MICINN under grants PID2021-123012, PID2021-128989 PID2022-141079, SEV-2016-0588, CEX2020-001058-M and CEX2020-001007-S, some of which include ERDF funds from the European Union. IFAE is partially funded by the CERCA program of the Generalitat de Catalunya.

We  acknowledge support from the Brazilian Instituto Nacional de Ci\^encia e Tecnologia (INCT) do e-Universo (CNPq grant 465376/2014-2).

\section*{Appendix}

In this section, we present Table \ref{table_a_candidates}, which summarizes key information about the strong lensing candidates in the A category. Additionally, we provide a complete table with this information for all subjects inspected by our experts, available in a machine-readable format accompanying this publication.

\addtocounter{table}{-1}
\suppressfloats[t]  
\begin{longtable*}{c c c c c p{5cm}} 
\caption{Catalog of strong lensing candidates in our category A}

\label{table_a_candidates} \\

\hline
Name              & ID           & RA (deg)        & Dec (deg)       & Exp. score & References \\ 
\hline
\endfirsthead

\hline
Name              & ID           & RA         & DEC        & Exp. score & References \\ 
\hline
\endhead

\hline
\endfoot

DESJ000316-334804&1032630238&0.818284&-33.801233&2.68&\cite{2019ApJS..243...17J, 2019MNRAS.484.5330J} \\
DESJ000451-010318&1040460697&1.21556&-1.055098&2.55&\cite{2017ApJS..232...15D, 2020MNRAS.495.1291J, 2022AA...668A..73R} \\
DESJ000816-323714&1040688993&2.067008&-32.620741&2.42&\cite{2019AA...632A..56K, storfer2023newstronggravitationallenses} \\
DESJ001110-323704&1046194283&2.793084&-32.617921&2.29&This work \\
DESJ001542-463611&1054556683&3.928374&-46.603077&3.0&\cite{2019ApJS..243...17J, 2022AA...668A..73R} \\
DESJ002510-494626&1065068829&6.294487&-49.774042&2.72&\cite{2022ApJS..259...27O, 2022AA...668A..73R, 2022ApJ...932..107S, storfer2023newstronggravitationallenses} \\
DESJ003104-440300&1075990880&7.770369&-44.050078&2.26&\cite{2017ApJS..232...15D, 2022AA...668A..73R} \\
DESJ003119-642037&1080776069&7.832104&-64.343818&2.41&\cite{2019ApJS..243...17J} \\
DESJ003507-252658&1079483541&8.7806&-25.449639&2.83&\cite{2019ApJS..243...17J, 2022ApJS..259...27O, 2022AA...668A..73R} \\
DESJ003727-413150&1083587736&9.362854&-41.530574&2.55&\cite{2017ApJS..232...15D, 2019ApJS..243...17J, 2022AA...668A..73R} \\
DESJ004109-004349&1088494058&10.287583&-0.730289&2.53&\cite{2019ApJS..243...17J, 2020ApJ...894...78H} \\
DESJ004144-233905&1093364977&10.434067&-23.651545&2.5&\cite{2021ApJ...909...27H} \\
DESJ004150-515750&1091516325&10.461274&-51.963971&2.3&\cite{2022ApJ...932..107S, storfer2023newstronggravitationallenses} \\
DESJ004257-371858&1093259184&10.738893&-37.31623&2.71&\cite{2019ApJS..243...17J} \\
DESJ004958+022714&1102641485&12.493282&2.454043&2.74&\cite{2022ApJ...932..107S, storfer2023newstronggravitationallenses} \\
DESJ005736-484814&1120481923&14.403741&-48.803999&2.68&\cite{2017ApJS..232...15D, 2022ApJS..259...27O} \\
DESJ005948-244220&1122534660&14.952536&-24.705638&2.58&\cite{2021ApJ...909...27H} \\
DESJ010519+014456&1135365822&16.331889&1.749025&2.87&\cite{2019ApJS..243...17J, 2020ApJ...894...78H} \\
DESJ010606-310437&1133540843&16.525913&-31.077073&2.27&\cite{2020ApJ...899...30L, 2021ApJ...909...27H, 2022ApJS..259...27O} \\
DESJ010659-443201&1136639478&16.746433&-44.533746&2.84&\cite{2019ApJS..243...17J, 2022AA...668A..73R} \\
DESJ010704-312840&1139453426&16.770516&-31.478048&2.83&\cite{2020ApJ...899...30L, 2021ApJ...909...27H, 2022ApJS..259...27O} \\
DESJ011333-381312&1147504166&18.389683&-38.220225&2.57&\cite{2022AA...668A..73R, 2022ApJ...932..107S, storfer2023newstronggravitationallenses} \\
DESJ011408-361313&1151529929&18.534964&-36.220442&2.28&\cite{2019ApJS..243...17J} \\
DESJ011646-243702&1153333586&19.194995&-24.61728&2.39&\cite{2019ApJS..243...17J, 2022AA...668A..73R} \\
DESJ011756-242824&1154333895&19.48702&-24.473544&2.27&\cite{2019ApJS..243...17J} \\
DESJ011758-052717&1159257785&19.494912&-5.45495&2.37&\cite{2019ApJS..243...17J, 2020ApJ...894...78H, 2022AA...668A..73R} \\
DESJ011842-615613&1162332513&19.677498&-61.937023&2.39&\cite{2019ApJS..243...17J, storfer2023newstronggravitationallenses} \\
DESJ012025-182001&1164169304&20.107368&-18.333831&2.26&\cite{2019ApJS..243...17J, 2019MNRAS.484.5330J} \\
DESJ012042-514353&1163516688&20.176065&-51.731425&3.0&\cite{2017ApJS..232...15D, 2019ApJS..243...17J, 2020MNRAS.494.1308N, 2022AA...668A..73R} \\
DESJ012431-520703&1170335487&21.130103&-52.117501&2.48&\cite{2019ApJS..243...17J} \\
DESJ012533-414217&1172896430&21.389999&-41.704998&2.83&\cite{2017ApJS..232...15D, 2019ApJS..243...17J} \\
DESJ012611-131642&1173896612&21.549582&-13.278375&2.85&\cite{2019ApJS..243...17J} \\
DESJ012753-453233&1178407449&21.971634&-45.542754&2.53&\cite{2019ApJS..243...17J} \\
DESJ012933-150634&1176674664&22.388727&-15.109646&2.59&\cite{2021ApJ...909...27H, 2022AA...668A..73R} \\
DESJ013002-374457&1179738251&22.512071&-37.749394&2.83&\cite{2019ApJS..243...17J, 2022ApJS..259...27O, 2022AA...668A..73R, storfer2023newstronggravitationallenses} \\
DESJ013026-152012&1177839322&22.610422&-15.336928&2.46&\cite{2019ApJS..243...17J, 2020ApJ...894...78H, 2022AA...668A..73R, storfer2023newstronggravitationallenses} \\
DESJ013050-160008&1178164184&22.710647&-16.002424&2.3&\cite{2019ApJS..243...17J, 2020ApJ...894...78H} \\
DESJ013322-125201&1300215437&23.342167&-12.867052&2.37&\cite{2019ApJS..243...17J} \\
DESJ013354-643412&1186271916&23.47771&-64.57026&2.86&\cite{2019ApJS..243...17J, 2022ApJS..259...27O} \\
DESJ013522-423223&1305035224&23.845168&-42.539875&2.56&\cite{2017ApJS..232...15D, 2019ApJS..243...17J, 2022AA...668A..73R} \\
DESJ013542-203335&1301081250&23.928359&-20.559886&2.25&\cite{2019ApJS..243...17J, 2022ApJS..259...27O, 2022AA...668A..73R} \\
DESJ013822-284408&1309246601&24.595731&-28.735568&2.38&\cite{2019ApJS..243...17J, 2021ApJ...909...27H, 2022ApJS..259...27O, 2022AA...668A..73R} \\
DESJ014134-404033&1307017630&25.391709&-40.675957&2.43&\cite{2019ApJS..243...17J} \\
DESJ014234-164817&1301526898&25.64572&-16.804905&3.0&\cite{2019ApJS..243...17J} \\
DESJ014252-183115&1620013669&25.720342&-18.521055&2.87&\cite{2019ApJS..243...17J, 2019MNRAS.484.5330J, 2022AA...668A..73R} \\
DESJ014326-085021&1194748640&25.862212&-8.839277&2.68&\cite{2019ApJS..243...17J, 2019MNRAS.484.5330J, 2022ApJS..259...27O, 2022AA...668A..73R} \\
DESJ014433-114209&1195735476&26.138987&-11.70261&2.38&\cite{2019ApJS..243...17J, 2020ApJ...894...78H, 2022AA...668A..73R} \\
DESJ014546-354127&1196484135&26.44499&-35.690953&2.87&\cite{2019ApJS..243...17J, 2022ApJS..259...27O, 2022AA...668A..73R} \\
DESJ014556+040228&1194195904&26.484828&4.041358&2.59&\cite{2019ApJS..243...17J, 2020ApJ...894...78H, 2022AA...668A..73R} \\
DESJ014655-092959&1311861257&26.731673&-9.499843&2.51&This work \\
DESJ014839-442555&1205436665&27.16306&-44.43212&2.68&\cite{2022ApJ...932..107S} \\
DESJ014908-313738&1200974180&27.283341&-31.627336&2.44&\cite{2019ApJS..243...17J, 2021ApJ...923...16L} \\
DESJ015225-570359&1207637405&28.108001&-57.066558&2.68&\cite{storfer2023newstronggravitationallenses} \\
DESJ015737-122138&1223104292&29.404393&-12.36064&2.31&\cite{2021ApJ...909...27H} \\
DESJ015824-003959&1224505431&29.603206&-0.666636&2.7&\cite{2019ApJS..243...17J, 2020ApJ...894...78H, 2022ApJS..259...27O, 2022AA...662A...4S} \\
DESJ020107-155117&1225062343&30.283231&-15.854756&2.61&\cite{2019ApJS..243...17J, 2020ApJ...894...78H, 2022AA...668A..73R} \\
DESJ020134-603640&1227758667&30.392541&-60.611218&2.83&\cite{2022ApJ...932..107S} \\
DESJ020144-273942&1229530106&30.436133&-27.661776&2.87&\cite{2019ApJS..243...17J, 2021ApJ...909...27H} \\
DESJ020304-233802&1235391858&30.766741&-23.634049&2.83&\cite{2019ApJS..243...17J, 2019MNRAS.484.5330J, 2020AA...644A.163C, 2022ApJS..259...27O, 2022AA...668A..73R} \\
DESJ020505-403830&1621099632&31.27143&-40.641747&2.83&\cite{2017ApJS..232...15D, 2019ApJS..243...17J, 2020MNRAS.494.1308N, 2022AA...668A..73R} \\
DESJ020526-353947&1234418937&31.358766&-35.66318&3.0&\cite{2019ApJS..243...17J, 2021ApJ...923...16L, 2022ApJS..259...27O} \\
DESJ020611-372938&1240229572&31.547501&-37.494102&2.5&\cite{2022ApJ...932..107S, storfer2023newstronggravitationallenses} \\
DESJ020706-272644&1243683190&31.777784&-27.445807&3.0&\cite{2019ApJS..243...17J, 2021ApJ...909...27H, 2021ApJ...923...16L, 2022ApJS..259...27O} \\
DESJ020954-354757&1248376493&32.477616&-35.79924&2.42&This work \\
DESJ021100-193810&1247965442&32.752503&-19.63615&2.6&\cite{2020AA...644A.163C} \\
DESJ021514-290925&1255097893&33.809596&-29.157161&2.85&\cite{2019ApJS..243...17J, 2019AA...632A..56K, 2021ApJ...909...27H} \\
DESJ021744-215834&1260779971&34.43687&-21.976185&2.26&\cite{storfer2023newstronggravitationallenses} \\
DESJ022140-021020&1272147895&35.417247&-2.172297&2.26&\cite{2018PASJ...70S..29S, 2019ApJS..243...17J, 2021AA...653L...6C, 2022AA...662A...4S} \\
DESJ022148-642642&1264873479&35.453169&-64.445139&2.38&\cite{2022AA...668A..73R, 2022ApJ...932..107S, storfer2023newstronggravitationallenses} \\
DESJ022310-224817&1270874866&35.794587&-22.804847&2.7&\cite{2021ApJ...909...27H, 2022AA...668A..73R} \\
DESJ022809-125252&1282567543&37.03776&-12.881181&2.26&\cite{2019ApJS..243...17J, 2020ApJ...894...78H} \\
DESJ022930-290816&1286124138&37.379004&-29.137926&2.58&\cite{2019ApJS..243...17J} \\
DESJ023249-032326&1287909989&38.207791&-3.390566&3.0&\cite{2017ApJS..232...15D, 2019ApJS..243...17J, 2020ApJ...894...78H, 2020MNRAS.495.1291J, 2021AA...653L...6C, 2022AA...662A...4S} \\
DESJ023341-344759&1294274501&38.422054&-34.79983&2.87&\cite{2021ApJ...923...16L, 2022ApJ...932..107S} \\
DESJ023906-204718&1321063152&39.776981&-20.788368&2.29&\cite{2019ApJS..243...17J} \\
DESJ024328-214202&1326680132&40.866699&-21.700556&2.32&\cite{2019ApJS..243...17J} \\
DESJ024604-060739&1331836609&41.520481&-6.12752&2.68&\cite{2020ApJ...894...78H} \\
DESJ024809-395548&1334070236&42.039764&-39.930101&2.28&\cite{2019ApJS..243...17J, 2022ApJS..259...27O, 2022AA...668A..73R, storfer2023newstronggravitationallenses} \\
DESJ024857-605403&1335051094&42.240122&-60.900967&2.29&\cite{2019ApJS..243...17J} \\
DESJ025052-552411&1342755111&42.717888&-55.403254&2.27&\cite{2019ApJS..243...17J, 2022AA...668A..73R} \\
DESJ025623-270718&1358779892&44.097386&-27.121845&3.0&\cite{2019ApJS..243...17J} \\
DESJ030639-273626&1368575074&46.665437&-27.607478&2.27&\cite{2019MNRAS.484.3879P, 2021ApJ...909...27H} \\
DESJ031127-423219&1380201297&47.863261&-42.538645&2.53&\cite{2019ApJS..243...17J, 2022ApJS..259...27O} \\
DESJ031638-223633&1441805121&49.161822&-22.609255&2.53&\cite{2019ApJS..243...17J, 2022AA...668A..73R} \\
DESJ031904-492143&1445784294&49.768511&-49.362018&2.25&\cite{2022AA...668A..73R, 2022ApJ...932..107S, storfer2023newstronggravitationallenses} \\
DESJ031906-273941&1444249631&49.778598&-27.66145&2.58&\cite{2021ApJ...909...27H} \\
DESJ031941-173404&1447194025&49.922007&-17.567922&2.83&\cite{2019ApJS..243...17J} \\
DESJ032125-153003&1449752819&50.356635&-15.501013&2.34&\cite{2019ApJS..243...17J} \\
DESJ032216-523440&1385113597&50.568423&-52.577903&2.87&\cite{2017ApJS..232...15D, 2019ApJS..243...17J, 2020MNRAS.494.1308N, 2022AA...668A..73R} \\
DESJ032449-102053&1382896242&51.205155&-10.348143&2.72&\cite{2020AA...644A.163C, 2021ApJ...909...27H} \\
DESJ032711-324634&1386748971&51.797337&-32.776162&2.29&\cite{2019ApJS..243...17J, 2019MNRAS.484.5330J, 2022AA...668A..73R} \\
DESJ033056-522815&1395105474&52.734562&-52.47084&2.7&This work \\
DESJ033143-612315&1397305017&52.932554&-61.3875&2.56&\cite{2022AA...668A..73R, 2022ApJ...932..107S, storfer2023newstronggravitationallenses} \\
DESJ033717-315213&1400483784&54.321876&-31.870435&3.0&\cite{2019ApJS..243...17J, 2022AA...668A..73R} \\
DESJ034130-513045&1410370257&55.378418&-51.51253&2.45&\cite{2017ApJS..232...15D, 2019ApJS..243...17J, 2022ApJS..259...27O, 2022AA...668A..73R} \\
DESJ034311-282840&1409295264&55.797722&-28.477789&2.42&\cite{2021ApJ...909...27H, 2023arXiv230111080H, 2022arXiv220607714L} \\
DESJ034435-444744&1413196615&56.146233&-44.795611&2.83&\cite{2019ApJS..243...17J} \\
DESJ034744-245431&1414531628&56.935597&-24.908735&2.42&\cite{2019ApJS..243...17J, 2019MNRAS.484.5330J, 2022AA...668A..73R} \\
DESJ035242-382544&1425127607&58.17678&-38.429166&3.0&\cite{2019ApJS..243...17J, 2022ApJS..259...27O, 2022AA...668A..73R} \\
DESJ035337-402417&1423677815&58.405967&-40.404958&2.27&\cite{2017ApJS..232...15D, 2022AA...668A..73R} \\
DESJ035346-170639&1423965682&58.442713&-17.110922&2.83&\cite{2019ApJS..243...17J, 2020AA...644A.163C} \\
DESJ035418-160952&1425763095&58.576161&-16.164526&2.87&\cite{2019ApJS..243...17J, 2019MNRAS.484.5330J, 2022AA...668A..73R} \\
DESJ035510-183134&1427368966&58.792571&-18.52621&2.29&\cite{2021ApJ...909...27H} \\
DESJ035649-240841&1429256248&59.204423&-24.144758&2.39&\cite{2019ApJS..243...17J, 2022ApJS..259...27O, 2022AA...668A..73R} \\
DESJ035713-475652&1432852433&59.306552&-47.948003&2.72&This work \\
DESJ040624-264624&1453825350&61.601641&-26.773339&3.0&\cite{2019ApJS..243...17J} \\
DESJ040715-571303&1458095582&61.815154&-57.217543&2.83&\cite{2019ApJS..243...17J, storfer2023newstronggravitationallenses} \\
DESJ040822-532714&1454069093&62.094448&-53.453918&2.83&\cite{2017ApJS..232...15D, 2022AA...668A..73R} \\
DESJ041112-541320&1462051999&62.802003&-54.222443&2.26&\cite{2022AA...668A..73R} \\
DESJ041242-264632&1616630786&63.178811&-26.775662&2.87&\cite{2019ApJS..243...17J} \\
DESJ041644-552500&1466200592&64.186786&-55.416745&2.55&\cite{2017ApJS..232...15D, 2019ApJS..243...17J, 2022ApJS..259...27O} \\
DESJ041809-545734&1466401262&64.541216&-54.959696&2.59&\cite{2017ApJS..232...15D, 2019ApJS..243...17J, 2020MNRAS.494.1308N, 2022AA...668A..73R} \\
DESJ042048-563046&1468847242&65.201028&-56.51302&2.54&\cite{storfer2023newstronggravitationallenses} \\
DESJ042218-213245&1470342587&65.575916&-21.546111&2.85&\cite{2019ApJS..243...17J, 2019MNRAS.484.5330J, 2022AA...668A..73R} \\
DESJ042723-290743&1481457559&66.848597&-29.128656&2.83&This work \\
DESJ043303-271423&1491366260&68.26478&-27.239827&3.0&\cite{2019ApJS..243...17J} \\
DESJ043806-322852&1493630017&69.525777&-32.481174&2.69&\cite{2019ApJS..243...17J} \\
DESJ044747-304630&1505802492&71.94812&-30.775079&2.41&This work \\
DESJ044805-580721&1505900316&72.022081&-58.122569&2.27&\cite{2017ApJS..232...15D, 2019ApJS..243...17J, 2022AA...668A..73R} \\
DESJ045008-571519&1510652342&72.536758&-57.255536&3.0&\cite{2017ApJS..232...15D, 2019ApJS..243...17J, 2020MNRAS.494.1308N} \\
DESJ045352-502234&1516572409&73.470762&-50.376362&2.56&\cite{2022AA...668A..73R, 2022ApJ...932..107S, storfer2023newstronggravitationallenses} \\
DESJ045951-304324&1523439802&74.964266&-30.723591&2.55&\cite{2021ApJ...909...27H, 2022AA...668A..73R} \\
DESJ050849-214430&1536678003&77.205297&-21.741881&2.37&\cite{2019ApJS..243...17J, 2019MNRAS.484.5330J} \\
DESJ051314-523020&1538255919&78.3096&-52.505683&2.55&\cite{2022AA...668A..73R, storfer2023newstronggravitationallenses} \\
DESJ051325-305035&1540902420&78.356147&-30.843306&3.0&\cite{2021ApJ...909...27H, 2022ApJS..259...27O} \\
DESJ052700-185805&1559106390&81.753525&-18.96833&2.28&This work \\
DESJ053804-473513&1569046306&84.51926&-47.587146&2.5&\cite{2017ApJS..232...15D, 2019ApJS..243...17J, 2020MNRAS.494.1308N, 2022AA...668A..73R} \\
DESJ054735-600442&1584250649&86.897808&-60.078518&2.37&\cite{2017ApJS..232...15D, 2022AA...668A..73R} \\
DESJ054848-421254&1588997152&87.200728&-42.215103&2.3&\cite{2022AA...668A..73R, 2022ApJ...932..107S, storfer2023newstronggravitationallenses} \\
DESJ055734-415950&1591517201&89.393433&-41.997283&2.43&\cite{2019ApJS..243...17J, storfer2023newstronggravitationallenses} \\
DESJ060239-465307&1601363180&90.662686&-46.885496&2.25&\cite{2017ApJS..232...15D} \\
DESJ060246-452443&1597714254&90.695238&-45.411976&2.87&\cite{2017ApJS..232...15D, 2019ApJS..243...17J, 2020MNRAS.494.1308N} \\
DESJ060443-414837&1599167680&91.180479&-41.810311&2.27&\cite{storfer2023newstronggravitationallenses} \\
DESJ060653-585843&1604164262&91.721436&-58.978777&2.87&\cite{2022AA...668A..73R, 2022ApJ...932..107S, storfer2023newstronggravitationallenses} \\
DESJ062415-470942&1610316947&96.065874&-47.161667&2.29&\cite{2019ApJS..243...17J, 2019MNRAS.484.5330J, 2022AA...668A..73R} \\
DESJ203911-545942&895557306&309.797344&-54.995216&2.28&\cite{2017ApJS..232...15D, 2019ApJS..243...17J, 2022AA...668A..73R} \\
DESJ211005-563930&909363984&317.522564&-56.658496&2.27&\cite{2017ApJS..232...15D, 2019ApJS..243...17J} \\
DESJ211627-594701&913936319&319.113845&-59.783864&2.87&\cite{2019ApJS..243...17J} \\
DESJ212251-005949&918844670&320.716645&-0.997037&2.38&\cite{2017ApJS..232...15D, 2022ApJS..259...27O} \\
DESJ212427-612510&919882930&321.113269&-61.419484&3.0&\cite{2022ApJS..259...27O, storfer2023newstronggravitationallenses} \\
DESJ213054-485714&925546537&322.727846&-48.954092&2.56&\cite{2022ApJ...932..107S} \\
DESJ221912-434835&958711616&334.801696&-43.809809&2.83&\cite{2019ApJS..243...17J, 2022ApJS..259...27O, 2022AA...668A..73R} \\
DESJ224434-590910&974192911&341.144978&-59.152836&2.26&\cite{2022ApJ...932..107S, storfer2023newstronggravitationallenses} \\
DESJ225403-405549&981794120&343.512752&-40.930347&2.42&\cite{2017ApJS..232...15D, 2019ApJS..243...17J, 2022AA...668A..73R} \\
DESJ230521-000211&995519978&346.34027&-0.036597&2.86&\cite{2018ApJ...867..107W, 2019ApJS..243...17J, 2021AA...653L...6C, 2022AA...662A...4S} \\
DESJ232128-463049&1005730815&350.368333&-46.513736&2.83&\cite{2017ApJS..232...15D, 2019ApJS..243...17J, 2020MNRAS.494.1308N, 2022AA...668A..73R} \\
DESJ233459-640407&1015482810&353.746734&-64.068626&3.0&\cite{2019ApJS..243...17J, 2022AA...668A..73R} \\
DESJ233551-515217&1015396778&353.966419&-51.871635&2.68&\cite{2017ApJS..232...15D, 2019ApJS..243...17J, 2022ApJS..259...27O, 2022AA...668A..73R} \\
DESJ234930-511339&1027961497&357.375315&-51.22754&3.0&\cite{2017ApJS..232...15D, 2019ApJS..243...17J, 2020MNRAS.494.1308N, 2022ApJS..259...27O, 2022AA...668A..73R} \\
DESJ235717-571025&1031154124&359.323497&-57.173869&2.39&\cite{2019ApJS..243...17J} \\
DESJ235848-612558&1033271849&359.700322&-61.433041&2.83&\cite{2021ApJ...909...27H, 2022ApJS..259...27O} \\

\end{longtable*}
\onecolumngrid
\tablecomments{\raggedright The candidates are ordered by increasing RA. ID is the `COADD\_OBJECT\_ID', a unique object id for the object within the Y6 DES data release.}
\twocolumngrid
\suppressfloats[false]

\bibliography{sample631.bib}{}
\bibliographystyle{aasjournal}



\end{document}